\documentclass[10pt,twocolumn]{article}

\usepackage[utf8]{inputenc}
\usepackage[T1]{fontenc}

\usepackage{authblk}

\usepackage{graphicx}
\usepackage{amsmath,amsfonts,amssymb}
\usepackage{amsthm}
\usepackage{booktabs}
\usepackage{hyperref}
\usepackage{xcolor}
\usepackage{cite}

\usepackage{gensymb}
\usepackage{textcomp}
\usepackage{comment}
\usepackage{float}
\usepackage{subcaption}
\usepackage{multicol}
\usepackage{setspace}
\usepackage{tikz,lipsum}
\usetikzlibrary{positioning}

\usepackage{fancyvrb}
\usepackage{cleveref}
\usepackage{mathrsfs}
\usepackage{algorithmic}

\usepackage{mathtools}
\usepackage{mathpazo}

\def\BibTeX{{\rm B\kern-.05em{\sc i\kern-.025em b}\kern-.08em
    T\kern-.1667em\lower.7ex\hbox{E}\kern-.125emX}}
    
\title{Photonic Learning in Ultrafast Laser-Induced Complexity}

\author[1,2]{Fayad Ali Banna}
\author[1,2]{Eduardo Brandao}
\author[1]{Anthony Nakhoul}
\author[1,2,3]{Rémi Emonet}
\author[1,2]{Marc Sebban}
\author[1,*]{Jean-Philippe Colombier}

\affil[1]{Université Jean Monnet, CNRS, Laboratoire Hubert Curien UMR 5516, Institute of Optics Graduate School, Saint-Étienne F-42023, France}
\affil[2]{Inria team MALICE}
\affil[3]{Institut Universitaire de France (IUF), Paris, France}
\affil[*]{Corresponding author: jean.philippe.colombier@univ-st-etienne.fr}

\date{}

\begin{document}

\twocolumn[
\maketitle

\begin{abstract}
How can one design complex systems capable of learning for a given functionality? In the context of ultrafast laser-surface interaction, we unravel the nature of learning schemes tied to the emergence of complexity in dissipative structures. The progressive development of learning mechanisms, from direct information storage to the development of smart surfaces, originates from the network of curvatures formed in the unstable fluid under thermoconvective instability, which is subsequently quenched and resolidified. Under pulsed laser irradiation, non-equilibrium dynamics generate intricate nanoscale patterns, unveiling adaptive process mechanisms. We demonstrate that the imprints left by light act as a form of structural memory, encoding not only local effects directed by laser field polarization but also a cooperative strategy of reliefs that dynamically adjust surface morphology to optimize light capture. By investigating how apparent complexity and optical response are intricately intertwined, shaping one another, we establish a framework that draws parallels between material adaptation and learning dynamics observed in biological systems.
\end{abstract}

\vspace{1em}
]

%\section*{Introduction}
    
Intelligence extends beyond consciousness to the ability to interact and adapt to the subtleties of the surrounding universe, whether living or inanimate. This principle transcends the boundaries of biology to apply to material entities, such as surfaces excited by light. The interaction between a metallic surface and a pulsed energy flux, such as an ultrafast laser, offers a unique opportunity to understand the intrinsic mechanisms of learning, as the signature of the learning process can be sequentially imaged by the pulse-to-pulse irradiation. Norbert Wiener's cybernetic viewpoint reveals that machine intelligence is defined by its ability to efficiently process information to achieve specific goals\cite{wiener2019cybernetics}. This translates into a learning process where the machine adjusts its behaviors based on feedback from its environment, akin to the functioning of neural networks in artificial intelligence, which adjust their weights and connections based on training data. For photoexcited surfaces, the transient morphology reflects the nuanced response to electromagnetic fields on the surface reliefs, adapted or learned over time. 
This article explores how materials progressively self-organize in response to successive light flux, focusing on the evolution of surface asperities under repeated  pulses\cite{bonse2020laser,zhang2015coherence}, which, we show, can be seen as a \textit{learning mechanism}.

%Intelligent surfaces, 
Intelligent surfaces emerge from controlled nanoscale structuring, where collective and nonlocal interactions give rise to mesoscopic properties \cite{hansen2015continuum,rudenko2023light}. Shaped by near-field electromagnetic coupling\cite{rudenko2019self}, Van der Waals and intermolecular forces \cite{trice2008novel}, chemical bonding, and thermal fluctuations, they demonstrate how microscopic order drives distinctive macroscopic behaviors, enabling advanced photonic functionalities\cite{her1998microstructuring,gattass2008femtosecond} and adaptive materials \cite{sugioka2013ultrafast,ilday2017rich,makey2020universality}. Inspired by natural phenomena such as plant microstructures, surfaces with specific functionalities, such as self-cleaning or controlled adhesion, are designed using biomimetism\cite{zorba2008biomimetic,Stratakis2020Jul}. By integrating metaphotonic approaches to manipulate light at the nanoscale\cite{baev2015metaphotonics}, new perspectives emerge for creating intelligent, custom-made surfaces, with applications in optical sensors or 5D optical data storage\cite{shimotsuma2003self,yao2023materials,stoian2020advances}.

%Intelligent surfaces, at the intersection of condensed matter physics, nanotechnology, and materials science, explore the emergent properties resulting from organization and structuring at the nanoscale and microscopic level\cite{stoian2020advances,sugioka2013ultrafast}. At these scales, fundamental physical phenomena such as surface interactions, chemical bonds, Van der Waals and intermolecular forces, as well as thermal fluctuations and near-field electromagnetic interactions dominate\cite{rudenko2019self,rudenko2023light}, influencing the collective behavior of microscopic entities and their macroscopic reaction. For example, by controlling roughness and texture at the nanoscale, significant modifications can be made to the tribological properties of a surface, affecting friction and adhesion \cite{wakuda2003effect}. Similarly, 
 %Thus, intelligent surfaces highlight how an ordered microscopic structuring can lead to exceptional macroscopic behaviors, opening new possibilities for the development of photonic functionalities and adaptive materials\cite{ilday2017rich,makey2020universality}.

%Self-organized surfaces
Self-organized surfaces emerge from systems out of equilibrium in which entropy flux optimization favors the formation of dissipative structures at the nanoscale\cite{rudenko2023light,prigogine1963introduction}. Today, it is widely accepted that under the impact of ultra-short laser pulses, including thermoconvective instabilities, can be controlled reproducibly and predictively\cite{rudenko2020high,tsibidis2015ripples}. These systems exhibit remarkable collective effects at scales well below the wavelength of light, significantly altering local optical properties and capable of generating hybrid electromagnetic modes, localized surface plasmons, or local field enhancement effects\cite{rudenko2019self,perrakis2024impact}.

% changing surface with light is surprising
Under normal conditions, materials are not expected to modify their optical absorption properties simply due to steady illumination — a painted surface, for instance, does not darken when exposed to sunlight. However, in cases where the material’s albedo decreases upon exposure, a positive feedback loop can emerge: increased absorption alters surface properties in a way that further reduces reflectivity, amplifying energy uptake until the system reaches a critical threshold\cite{lu2018influence}.

% ... but life changes surface to optimize absorbtion all the time
Plants inherently regulate absorption up to a specific limit, beyond which instability ensues\cite{diaz2018mediterranean}. As they grow, they optimize light capture by adjusting leaf morphology, including size, shape, and orientation\cite{wang2022leaf}. However, excessive absorption can be detrimental. Mediterranean plants have evolved specialized strategies such as evaporative cooling and protective coatings, which contribute to their distinctive aromatic emissions\cite{penuelas2003bvocs}. They have developed selective light absorption mechanisms: leaf epidermal layers accumulate flavonoids in response to UV-B radiation, reducing epidermal transmittance and mitigating potential damage, while remaining sensitive to other wavelengths essential for photosynthesis\cite{teramura1994effects,middleton1994understanding}. Shaped by evolutionary pressures, these adaptations  dynamically regulate light absorption through microstructural variations, pigmentation adjustments, and controlled leaf orientation, preventing excessive energy uptake.\cite{lambers2008plant,chaves2009photosynthesis}.

% ... whereas innanimate things in general do not
In contrast, inanimate materials lack such adaptive mechanisms, making them highly susceptible to uncontrolled absorption feedback. A perfectly smooth metallic surface behaves as a mirror, absorbing energy in the visible-IR range only within a shallow skin depth\cite{maier2007plasmonics}. However, even nanoscale surface roughness disrupts this behavior, introducing spatial inhomogeneities in absorption and rendering it highly sensitive to the polarization of incident light\cite{ferry2011modeling,zhang2020laser}. Energy uptake on rough metallic surfaces arises from both radiative and non-radiative contributions: far-field scattering, governed by surface features relative to the incident wavelength, and near-field interactions, which involve localized plasmonic excitations and evanescent waves, leading to enhanced subwavelength-scale energy deposition\cite{novotny2012principles}. Under intense laser irradiation, this intricate interplay dictates the absorbed energy distribution, driving nanoscale pattern formation, localized heating, and surface restructuring\cite{rudenko2023light}.

% In this paper: innanimate surface learns to absorb
Informed by the reconstructed dynamics of laser-induced self-organized systems and guided by solutions of Maxwell equations, we develop a numerical approach that integrates machine learning to uncover the progressive stages of photonic information processing. We demonstrate that memory of previous irradiations is encoded as surface topography evolves into a complex 2D arrangement. As a result, optical coupling is optimized on textured surfaces, favoring specific patterns while filtering fluctuations to preserve relevant information such as light polarization and energy gradients.  %This reveals a fundamental learning mechanism, where information delivered by the light stimulus is stored and exploited to drive the structuring process. 
Despite the nonlinear dynamic, the learning trajectory follows a feedback-driven adaptation where hysteresis effects imprint a history-dependent evolution. We show that the emergence of surface complexity is directly tied to this adaptive learning dynamic, providing a key metric for understanding light-matter adaptation in laser patterning design.

\section*{Results}
\subsection*{Experimental self-organization}

\begin{figure*}[t]
\centering
\includegraphics[width=1\textwidth]{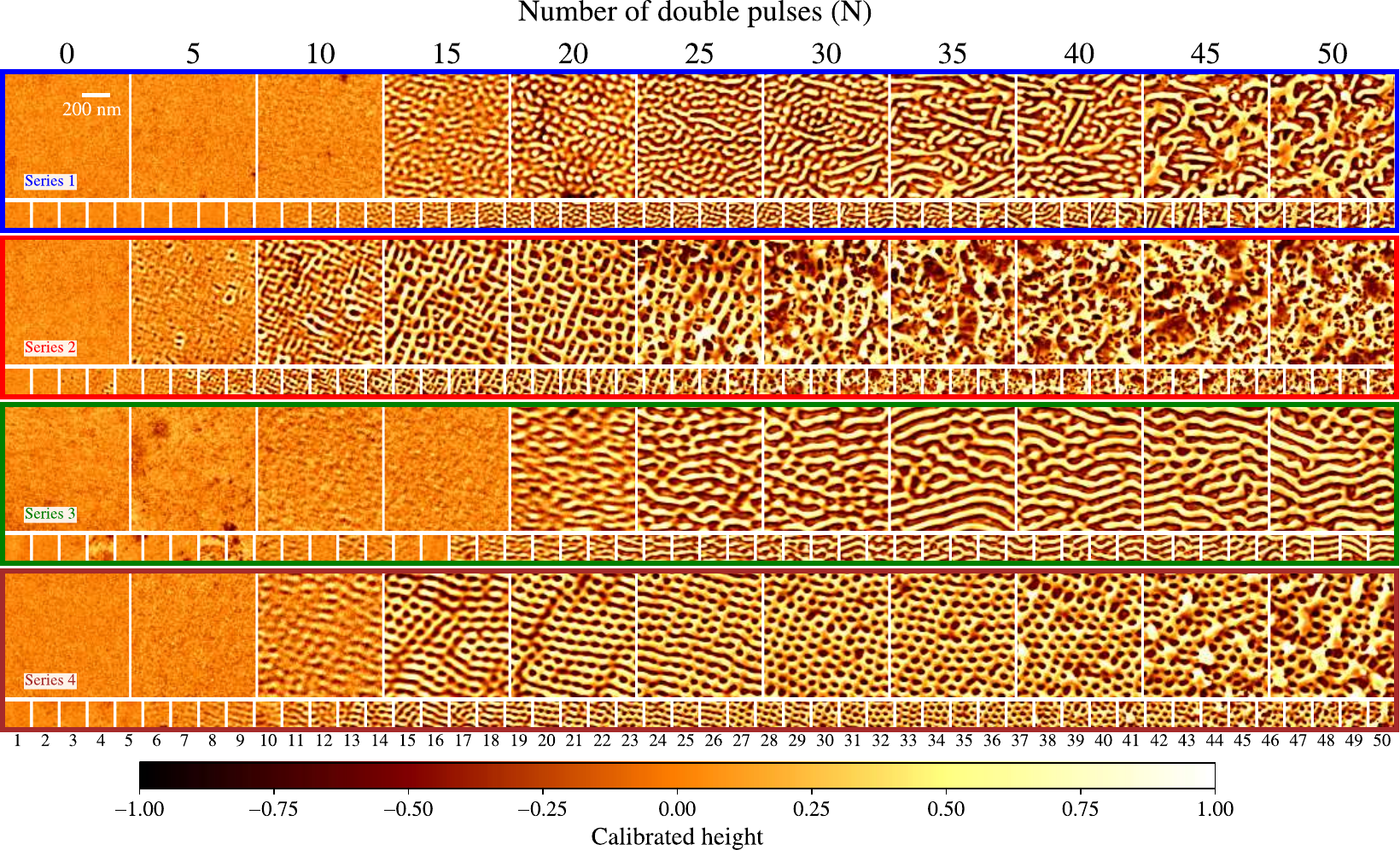}
\captionsetup{justification=justified, singlelinecheck=false}
\caption{\textbf{SEM images of nickel surfaces irradiated with cross-polarized double pulses $N$.} Each series represents a progressive formation of self-organized patterns for distinct combination of \(F\) and \(\Delta t\) : Series 1 correspond to \(F = 0.18\) J/cm² and \(\Delta t = 8\) ps, series 2  to \(F = 0.21\) J/cm² and \(\Delta t = 2\) ps, series 3 to \(F = 0.25\) J/cm² and \(\Delta t = 36\) ps and series 4 to \(F = 0.26\) J/cm² and \(\Delta t = 28\) ps. %The observed patterns in each series depend on these parameters and the number of double pulses. %As the number of double pulses increase, the patterns start to become chaotic.
}
\label{main-fig:figure1}
\end{figure*}
Optical feedback drives laser-induced self-organization. Under intense photoexcitation ($I\sim 10^{12}$ W.cm$^{-2}$), each pulse reshapes the surface, intensifying local heating and triggering Rayleigh–Bénard-Marangoni instabilities\cite{rudenko2020high}. Thermoconvection, governed by temperature gradients and rarefaction waves, gives rise to different patterns: hexagons, stripes, or peaks. During multipulse irradiation ($N$ pulses), cross-polarized double-pulse sequences reduce the anisotropy originating from single linear polarization, while the delay $\Delta t$ regulates the material's hydrodynamic response through an antagonistic light flux. A SEM (Scanning Electron Microscope) dataset was built by creating an impact with \(N\) double-pulses at 1 kHz, then translating the sample to create a new impact with \(N+1\) pulses. Since \((x,y)\) coordinates of \(u_{N}(x, y)\) and \(u_{N+1}(x, y)\) do not coincide, real-time local "pulse-to-pulse" evolution is absent. While structuring does not follow a simple additive model, global organization and features such as absorption properties remain robust for deriving the learning evolution of laser-induced patterning.

Fig.~\ref{main-fig:figure1} presents high-resolution SEM surface patterns, revealing high-resolution 2D nanostructures evolving with laser fluence \(F\), time delay \(\Delta t\), and pulse sequence \(N\). While 2D SEM images provide detailed surface topography, they inform intensity variations rather than absolute height. To extract depth information, we apply a grayscale conversion, assigning the brightest pixels (value$=255$) to the highest peaks ($+1$) and the darkest pixels ($0$) to the lowest valleys ($-1$).

In Series 1, nanopeaks form at \(F = 0.18\) J/cm$^{2}$ and \(\Delta t = 8\) ps, peaking in aspect ratio and concentration at \(N = 25\). Atomic force microscopy shows nanopeaks with approximately 100 nm height and 20 nm diameter (see Suppl. Mat.). For \(N < 5\), nanoroughness dominates, followed by convection cells (\(5 \leq N \leq 10\)). At \(N = 25\), the nanopeaks concentrate along the edges of the convection cell. Beyond \(N = 30\), nanopeaks merge into linear structures, and at \(N > 40\), dragon-like patterns emerge, indicating a transition to complex structures. This feedback process, in which each pulse interacts with the altered surface, drives self-organization through the dynamics of the molten material. In Series 2, nanowebs form at \(F = 0.21\) J/cm$^{2}$ and \(\Delta t = 2\) ps. For \(N > 10\), convection cells organize into nanowebs, reaching optimal structuring at \(N = 20\), but transitioning to chaotic structures for \(N > 25\). In Series 3, emerging at \(F = 0.25\) J/cm$^{2}$ and \(\Delta t = 36\) ps, convection cells at \(N = 10\) evolve into nanocavities (\(N > 20\)), forming labyrinthine structures at \(N \geq 35\). Similarly, in Series 4, shaping up at \(F = 0.26\) J/cm$^{2}$ and \(\Delta t = 28\) ps, hexagonal nanocavities emerge at \(N = 10\), becoming more organized for \(25 \leq N \leq 35\), while nanobumps form atop nanocavities for \(N > 45\), resulting in hierarchical structures.

\subsection*{Complexity drives absorption and absorption fuels complexity}Self-organized patterns arise from laser energy absorption, leading to localized heating and surface rearrangement via pressure gradients and surface tension forces. The resulting inhomogeneous energy distribution induces instabilities that trigger pattern formation \cite{rudenko2020high}. To analyze ultrafast light interactions with structured surfaces, we numerically solve Maxwell’s equations for varying polarization states, using the height map \( u_N(x,y) \) to uncover the interplay between energy absorption and self-organization.

\begin{figure*}[t]
    \centering
    \begin{tikzpicture}
        \node[anchor=north west] (img) at (0,0) 
            {\includegraphics[width=0.97\textwidth]{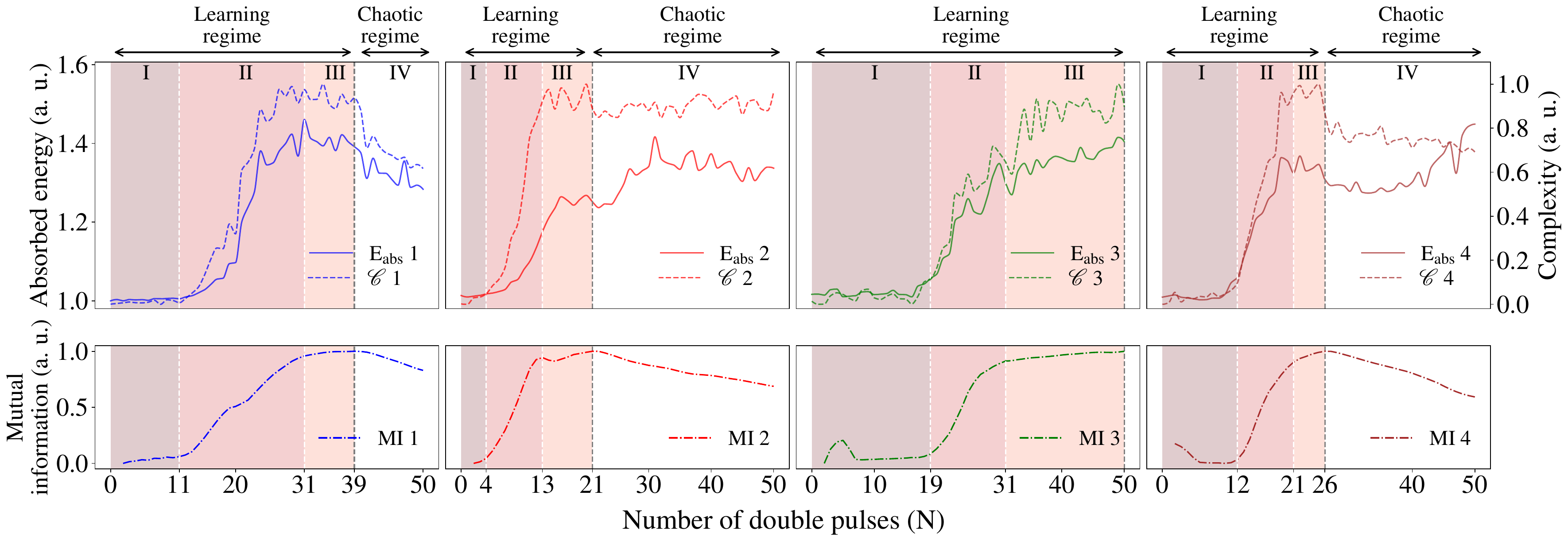}};
        
        \node[anchor=north west] at ([xshift=-4.5mm, yshift=-7mm]img.north west) {a};
        \node[anchor=north west] at ([xshift=-4.5mm, yshift=-37mm]img.north west) {b};
    \end{tikzpicture}
    \captionsetup{justification=justified, singlelinecheck=false}
    \caption{\textbf{Absorbed energy $E_{abs}$ correlated to complexity $\mathscr{C}$.} \textbf{(a)} $E_{abs}$ (solid line) and $\mathscr{C}$ (dashed line) as a function of $N$, ranging from 0 to 50 for the four series. The two variables go through 4 phases that reflect two regimes. The learning regime comprises : response phase (I), iterative learning phase (II) and memory stabilization phase (III). The chaotic regime is equivalent to the destruction phase (IV).
    \textbf{(b)} The mutual information $MI$ variation is the criteria to split $E_{abs}$ and $\mathscr{C}$ into 4 phases.}
    \label{fig:figure2}
\end{figure*}
Fig.~\ref{fig:figure2} strikingly reveals the strong correlation between absorbed energy and surface complexity for the four experimental series. %This demonstration is done by exploiting surfaces coming from four series of SEM images obtained experimentally by varying the number of double pulses between 0 and 50, for 4 combinations of laser fluence and intra-pulses delay. 
Fig.~\ref{fig:figure2}a displays the superimposed variation in absorbed energy and complexity as $N$ increases. Absorbed energy \( E_{\text{abs}} \) represents the average energy absorbed by the surface at the \( N \)-th laser pulse, considering polarization angles of \( \alpha=k\pi/6 \) for \( k = 0 \) to \( 5 \). This quantity is computed for a given polarization state and \( u_N(x,y) \) by integrating the energy absorption \( E_{\text{abs}} = \iint_{x,y} \varepsilon_{abs}(x,y) \,dx\,dy, \)  
where \( \varepsilon_{abs}(x,y) \) is the 2D energy absorption distribution integrated over depth.

A maximally complex system strikes a balance between coherent organization and randomness. In order to quantify this balance, we use \textit{Taylor complexity}\cite{PhD-Brandao-2024}, a statistical measure of dynamical complexity based on the López–Ruiz, Mancini, and Calbet (LMC) complexity \cite{lopez1995statistical}, defined as $C= \mathscr{H} \times \mathscr{D}$. In this expression, the Shannon entropy $\mathscr{H}$ quantifies the degree of randomness, while disequilibrium \( \mathscr{D} \) measures how far a system is from equilibrium, capturing the presence of structure or order. This formulation quantifies the degree of complexity that emerges when a system exhibits dissipative structures that are neither too ordered (low entropy, high disequilibrium) nor too random (high entropy, low disequilibrium).

We normalize absorbed energy by that of a flat surface to compare absorption propensity across different fluences, while  \( \mathscr{C} \) is also normalized to 1 to ensure both quantities are interpreted as relative increases or decreases. The evolution of absorbed energy and complexity unfolds through four distinct regimes indicated in Fig.~\ref{fig:figure2}, reflecting a dynamic feedback-driven adaptation process governed by the progressive absorption of ultrafast pulses. Initially, during the \emph{response} phase (I), the system undergoes an exploratory stage where local topographical changes emerge in response to the stimulus, mirroring an early learning process. At this stage, hydrodynamic instabilities begin to develop but remain transient, with \( E_{\text{abs}} \) and \( \mathscr{C} \) relatively stable. A sharp increase in both quantities signals the transition to the \emph{iterative learning} phase (II). Active adaptation takes place and the surface refines its topography through self-organization as the absorbed energy and complexity evolve synchronously. This phase ends with the stabilization of hydrodynamic instability patterns, forming a strange attractor that becomes frozen in the surface morphology of the material. As the process progresses, the system enters the \emph{memory stabilization} phase (III), where its adaptive capacity stabilizes, indicating a form of memory consolidation. Complexity reaches a plateau and the material retains a structured response to light excitation, preserving the self-organized patterns. Then, prolonged exposure leads to the \emph{destruction} phase (IV), where excessive input disrupts the established organization, similarly to overlearning effects. In this final stage, absorbed energy and complexity decouple, with absorbed energy increasing while complexity declines. This destabilization erodes the previously frozen hydrodynamic patterns, leading to structural breakdown and the onset of chaotic behavior.

Fig.~\ref{fig:figure2}b illustrates the progression of mutual information $MI$ between the absorbed energy $E_{\text{abs}}$ and the complexity $\mathscr{C}$  with the number of double pulses $N$. $MI$ quantifies the dependency between two variables, capturing both linear and non-linear relationships. A high $MI$ indicates strong dependence between $E_{\text{abs}}$ and $\mathscr{C}$, while a low value suggests weak or no relationship. $MI$ serves as the criterion to frame \( E_{\text{abs}} \) and \( \mathscr{C} \) variation into the four previously mentioned phases. To compute $MI$, we consider data points $x_{N}=[E_{abs}^{N}, \mathscr{C}^{N}]^T$ where $x_{N}$ is a vector that represents the values of \( E_{\text{abs}} \) and \( \mathscr{C} \) after $N$ pulses. The computation considers points up to the given $N$, meaning that we incorporate all data points $x_{1}$, $x_{2}$, ..., $x_{N}$. Since at least three points are required for a meaningful estimate of mutual information, we start with $N=3$. To improve interpretability, we normalize $MI$ between 0 and 1, where 1 represents the maximum $MI$. Overall, the series exhibit a common behavior: for small $N$, $MI$ remains low and constant. As $N$ increases, $MI$ rises sharply until it reaches a plateau and a maximum, followed by a decline. However, in series 3, this drop does not occur and $MI$ continues to increase since there are no chaotic patterns. During the formation of well-organized structures \( E_{\text{abs}} \) and \( \mathscr{C} \) are initially correlated, but the emergence of chaotic patterns leads to their decorrelation, resulting in a reduction of $MI$. The value of $N$ at which this drop in $MI$ occurs marks the transition between the \emph{learning} and \emph{chaotic} regimes, as well as the boundary between \emph{memory stabilization} (III) and \emph{destruction} phases (IV). The beginning of sharp rise and the following plateau marks outline the other phases, allowing us to define the \emph{response} (I) and the \emph{iterative learning} (II) phases. 

\begin{figure*}[t]
\centering
\includegraphics[width=1\textwidth]{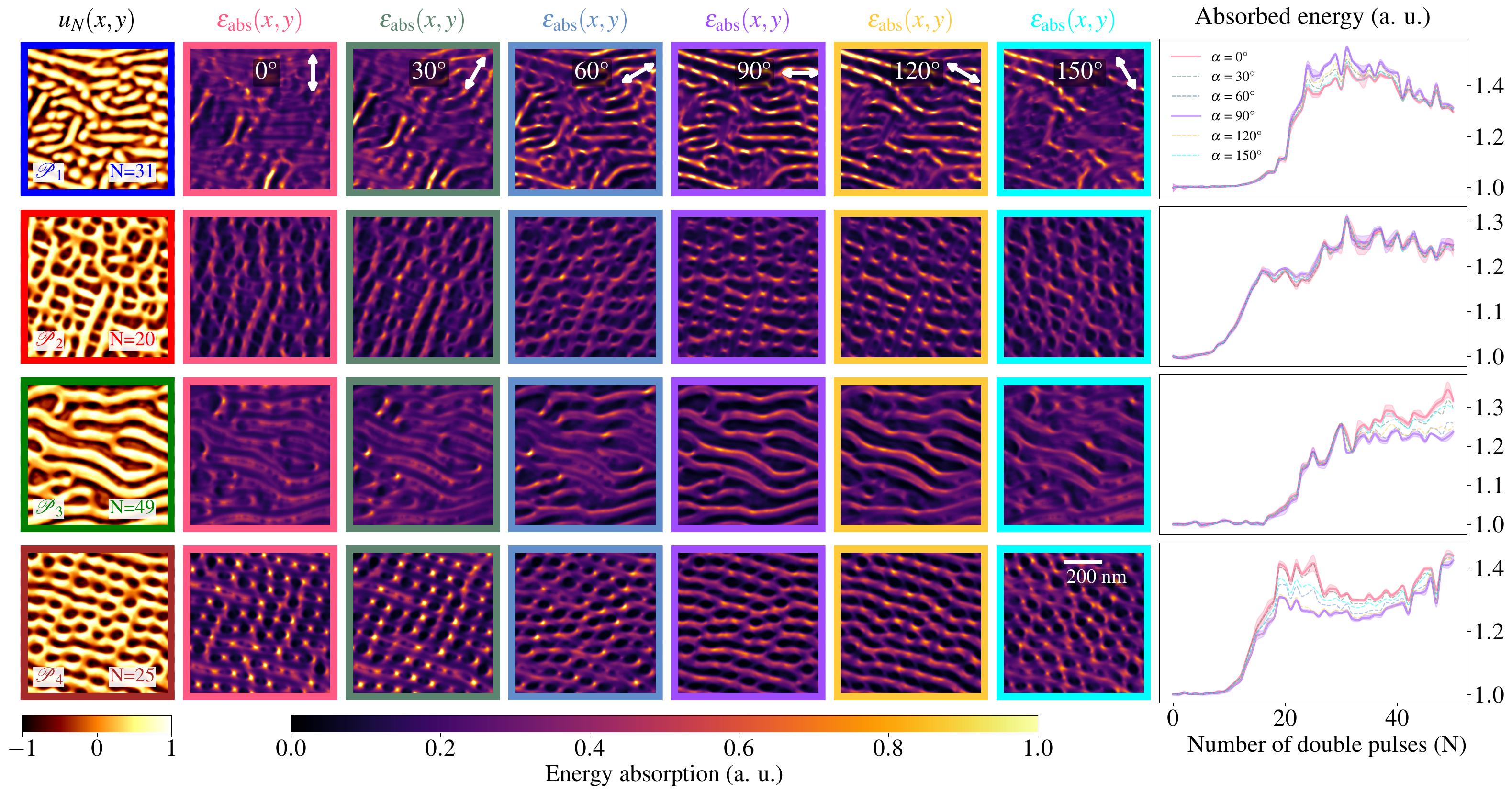}
\captionsetup{justification=justified, singlelinecheck=false}
\caption{\textbf{Effect of laser polarization angle $\alpha$ on absorbed energy $E_{\text{abs}}$.} Each row corresponds to a specific series \(i\) (\(i = 1\) to \(4\)). The columns, from left to right represent: the pattern height map \(u_{N}(x, y)\) corresponding to \(\mathscr{P}_{i}\), the pattern maximizing \(E_{\text{abs}}^{N+1}\) in series \(i\): \(\mathscr{P}_{1}\) (series 1, at \(N = 31\)), \(\mathscr{P}_{2}\)  (series 2, at \(N = 20\)), \(\mathscr{P}_{3}\) (series 3, at \(N = 49\)), and \(\mathscr{P}_{4}\) (series 4, at \(N = 25\)); the energy absorption from the subsequent pulse \( \varepsilon_{abs}^{N+1}(x,y) \) for \( \alpha=k\pi/6 \) for \( k = 0 \) to \( 5 \) and the last column is the integrated absorbed energy $E_{\text{abs}}^{N+1}=f(N,\alpha)$. The shaded region for \( \alpha=0 \) and \( \alpha=\pi/2 \)  represents a confidence band corresponding to the standard deviation of $E_{\text{abs}}^{N+1}$ calculated across different locations.  
}
\label{fig:figure3}
\end{figure*}

%To assess the specificity of the learning response as a function of polarization, Fig.~\ref{fig:figure3} summarizes the relationship between energy absorption and electromagnetic field polarization. 

%To differentiate the global response, which reflects the effect of the pattern on energy absorption, from the specific response, which corresponds to the effect of polarization, we analyze the relationship between energy absorption and the polarization of the electromagnetic field.

To distinguish the global response, which is the influence of patterns on energy absorption, from the specific response, which is the effect of polarization, we analyze how absorbed energy $E_{\text{abs}}$ varies with polarization $\alpha$. As shown in Fig.~\ref{fig:figure3}, four different self-organized patterns are used to generate six energy absorption maps per pattern, computed at polarization angles of \( \alpha=k\pi/6 \) for \( k = 0 \) to \( 5 \). Experimentally, self-organized structures are obtained using cross-polarized double pulses oriented at \( \alpha=0 \) and \( \alpha=\frac{\pi}{2} \), which we refer to as the experimental polarization angles, making them particularly significant. For a given series, after \( N \) double pulses, these maps show variations based on the alignment of the structures with the electric field $\vec{E}$. Structures parallel to the polarization direction absorb more energy, with peaks having the highest curvature maximizing absorption while valleys absorb less energy.

These observations are complemented by the evolution of absorbed energy as a function of the number of double pulses. For all considered polarization angles, the absorption curves follow a similar trend but remain distinctly separated across the range of double pulses $N$ (0 to 50). Initially, all curves overlap and remain constant as no clear structures affect energy absorption. As \( N \) increases, the curves dissociate, revealing distinct differences between polarization angles. This differentiation intensifies with the emergence of well-oriented structures. Notably, for the experimental polarization angles \( \alpha=0 \) and \( \alpha=\frac{\pi}{2} \), the curves of $E_{\text{abs}}$ consistently serve as the upper and lower bounds for the other angles. However, which one acts as the upper or lower bound varies between series. The light stimulus specificity determines how the structured surface enhances its absorption efficiency for the particular photoexcitation, with a given polarization angle \(\alpha\) leading to more targeted absorption patterns. Stimuli that are highly specific, aligning with the same orientation defined by \(\alpha\) as the previous \(N-1\) pulses, clearly enhance the system's response by matching the surface structures, while less specific stimuli with different \(\alpha\) angles result in a more uniform absorption distribution.

\begin{figure*}[t]
\centering
\includegraphics[width=1\textwidth]{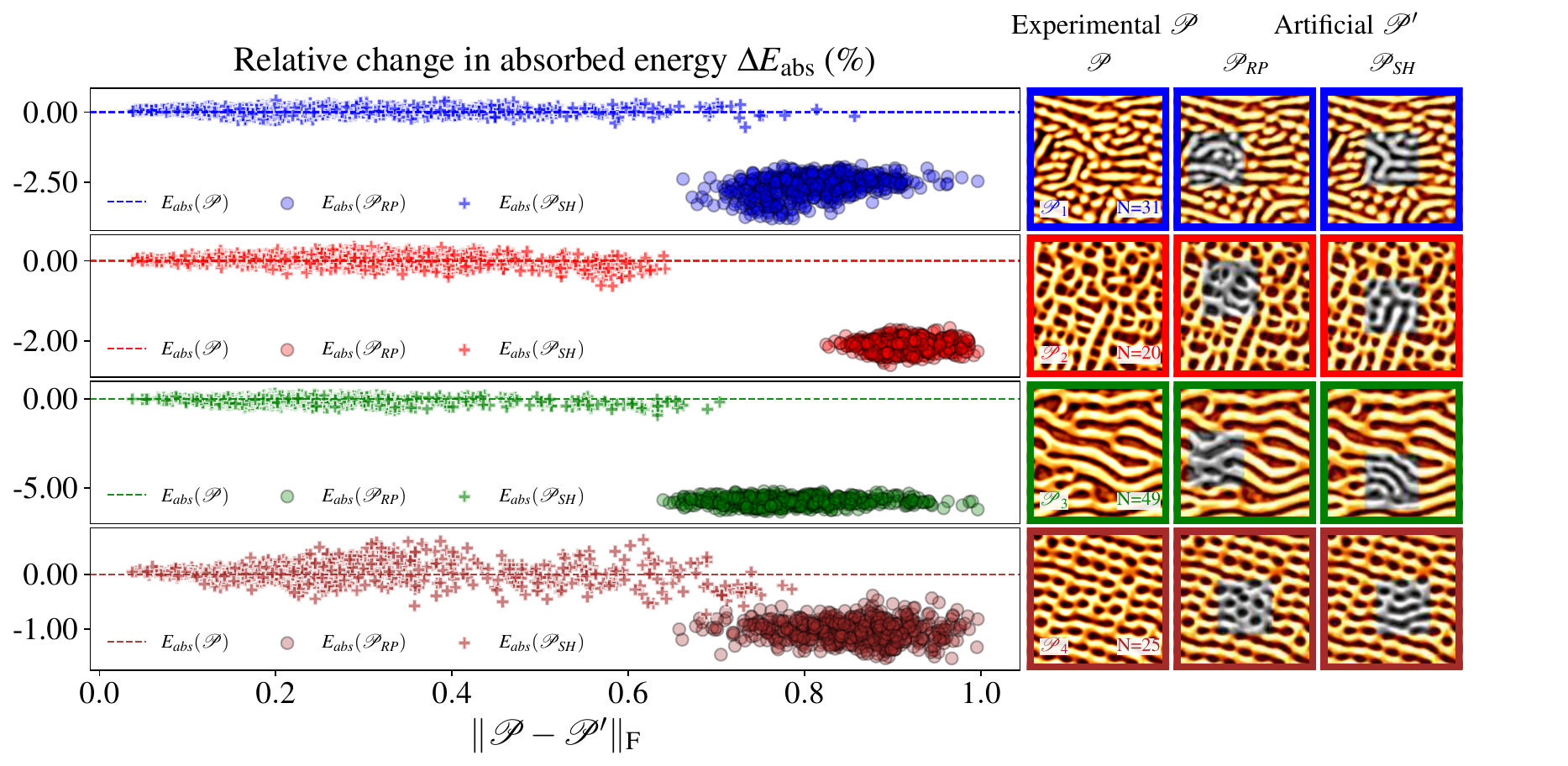}
%\caption{\textbf{Effect of modifying the structure of a pattern $\mathscr{P}$ maximizing energy absorption.} Absorbed energy $E_{\text{abs}}^{N+1}$ is represented as a function of the Frobenius norm $\|.\|_{F}$ between experimental and artificial surfaces. The first column (Experimental) displays surfaces obtained experimentally, the second (RP) corresponds to artificial surfaces created by randomly replacing a part of $\mathscr{P}$ by a patch from a related experimental surface. The last column (SH) displays artificial surfaces obtained by replacing a part of $\mathscr{P}$ by a patch produced by Swift-Hohenberg numerical solver. The highlighted gray areas indicate where a modification has been made.} 
\captionsetup{justification=justified, singlelinecheck=false}
\caption{\textbf{Effect of modifying the structure of a pattern $\mathscr{P}$ that maximizes energy absorption.} Each row correspond to a distinct series \(i\) (\(i = 1\) to \(4\)). On the right, the columns represent, from left to right :  the experimental pattern $\mathscr{P}$ that maximizes $E_{\text{abs}}^{N+1}$, the artificial pattern $\mathscr{P}_{RP}$ created by randomly replacing a part of $\mathscr{P}$ by a patch from the same series and the artificial pattern $\mathscr{P}_{SH}$ obtained by replacing a part of $\mathscr{P}$ by a patch produced by Swift-Hohenberg numerical solver. The highlighted gray areas indicate the modified areas relative to $\mathscr{P}$. On the left, the scatter plot shows, the relative change in absorbed energy $\Delta E_{\text{abs}}=100 \cdot \frac{E_{abs}(\mathscr{P}') - E_{abs}(\mathscr{P})}{E_{abs}(\mathscr{P})}$ as a function of the Frobenius norm $\|.\|_{F}$  between the experimental pattern $\mathscr{P}$ and its corresponding artificial version $\mathscr{P}'$ ($\mathscr{P}_{RP}$ or $\mathscr{P}_{SH}$) }
\label{fig:figure4}
\end{figure*}

To evaluate the effect of perturbing an experimental pattern $\mathscr{P}$, that maximizes the absorbed energy $E_{abs}$ for a given series i, we propose two distinct methods. Fig.~\ref{fig:figure4} summarizes the results, illustrating the relative change in absorbed energy $\Delta E_{abs}$ as a function of the Frobenius norm $\|.\|_{F}$ between $\mathscr{P}$ and its modified counterparts. The value of the Frobenius norm $\|\mathscr{P}-\mathscr{P}'\|_{F}$ is normalized to the range [0,1], ensuring a consistent scale for comparison. The value of $E_{abs}$ for $\mathscr{P}$, denoted as $E_{abs}(\mathscr{P})$ is indicated by the horizontal dashed line to compare $E_{abs}$ variation as $\mathscr{P}$ is modified. The first method modifies $\mathscr{P}$ by replacing a small patch with another patch randomly selected from one of the patterns in the same series, but within the learning regime. It creates artificial patterns $\mathscr{P}_{RP}$ that locally deviate from the self-organization laws. Computing $E_{abs}$ on $\mathscr{P}_{RP}$ produces scatter points $E_{abs}(\mathscr{P}_{RP})$ concentrated in a small region of the plot. This region consistently falls below the reference line, regardless of the similarity between the experimental and modified surfaces. In contrast, the second method replaces a portion of $\mathscr{P}$ with another patch generated using a Swift-Hohenberg (SH) solver. The SH equation models hydrodynamic fluctuations and generates pattern-like structures similar to the self-organized structures observed in our experiments. It is demonstrated that patterns formed by the process of self-organization can be approximately modeled by this equation \cite{brandao23_learn_compl_to_guide_light}. This method generates physically reasonable surfaces $\mathscr{P}_{SH}$ that preserve the dynamics of self-organization. Unlike the first method, the scatter points $E_{abs}(\mathscr{P}_{SH})$ are more widely distributed and fluctuate around the reference line. However, the variations remain minimal regardless of the similarity between the experimental and modified surfaces, indicating that self-organization fosters configurations that are resilient to perturbations. This vulnerability to perturbations highlights the fragility of self-organized patterns, demonstrating that they are finely tuned to optimize performance, and any deviation from this precise configuration weakens their effectiveness.

\section*{Learning strategy}
%Organization to absorb (Global introduction of the discussion)

\begin{figure}[t]
\centering
\includegraphics[width=0.5\textwidth]{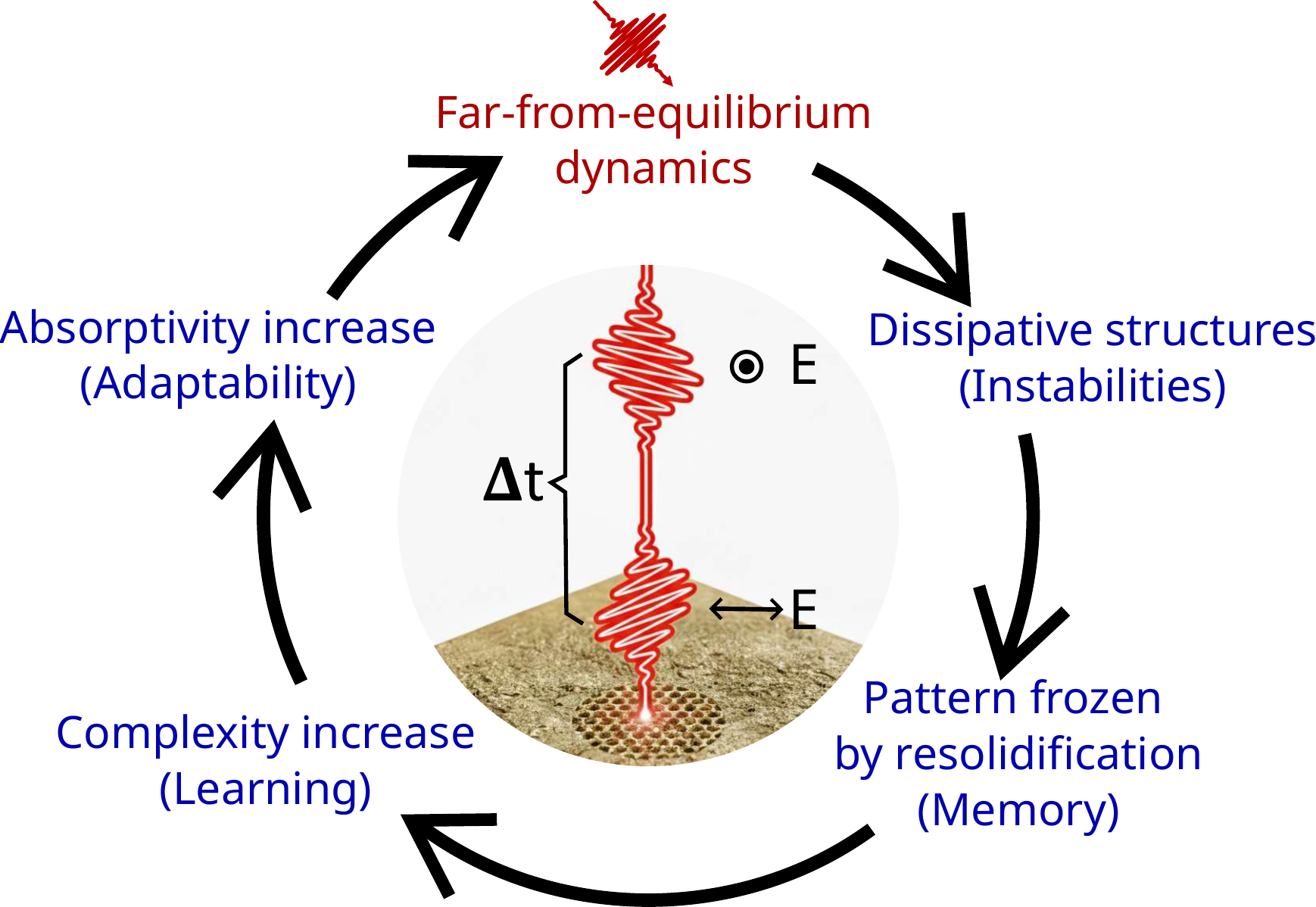}
\caption{\textbf{Schematic illustration of physical processes resulting from electromagnetic energy absorption.} $\leftrightarrow$ and $\odot$ denote two orthogonal, linear polarization states used successively, both aligned within the plane of the irradiated surface. Laser-induced surface self-organization originates from a dissipative instability in which morphologies that enhance optical absorption grow preferentially due to stronger energy dissipation and material transport. Repeated irradiation and resolidification progressively store this information in the surface morphology, leading to an effective learning process encoded in the surface complexity.}

\label{fig:learning_cycle}
\end{figure}

In this work, we define "learning" as the history-dependent evolution of surface morphology under repeated laser irradiation, through which the system is progressively guided toward patterns configurations that optimizes energy absorption. Under successive pulses, the morphology follows a well-defined trajectory in configuration space, and this trajectory exhibits two key properties that distinguish it from simple pattern formation.\\ 
First, the trajectory is history-dependent not merely because the surface topography at pulse $N$ serves as the initial condition for pulse $N+1$, but because the evolving morphology actively encodes information from previous electromagnetic interactions. In this way, the geometric state of the material constitutes a form of structural memory with the frozen topography shaping the optical response to the next pulses. In contrast, memoryless processes are those in which each pulse encounters statistically equivalent conditions, so no cumulative influence of past interactions exists.\\
Second, the trajectory shows directional evolution toward enhanced coupling. It is not a random walk in configuration space. Successive modifications of the surface progressively increase energy absorption and structural organization until the system approaches configurations that maximize light–matter interaction. These preferred states act as attractors toward which trajectories naturally converge under the influence of the repeated driving pulses.\\
The physical basis for this directed evolution lies in a feedback mechanism mediated by material restructuring. Each laser pulse deposits energy with a spatial distribution determined by the existing surface topography, leading to localized melting and thermoconvective transport. The resulting modifications of the surface geometry, in turn, alter the electromagnetic boundary conditions experienced by subsequent pulses. When these modifications enhance local absorption, they are thermally reinforced by later irradiation; conversely, reduced absorption leads to slower morphological growth.\\
This constitutes a nonlinear physical feedback loop in which surface morphology and optical response are locally coupled. Through repeated irradiation, the system progressively adapts its structure based on accumulated exposure history, optimizing absorption efficiency. This loop starting with each double pulses is schematically depicted in  Fig.~\ref{fig:learning_cycle}. The process follows a succession of stages: far-from-equilibrium dynamics, dissipative structures (instabilities), pattern frozen by resolidification (memory), complexity increase (learning), absorptivity increase (adaptability), and the return to far-from-equilibrium dynamics, completing the adaptive loop. It represents a process in which a system’s response evolves through experience to better match environmental demands, driven by the nonlinear dynamics of the system far from equilibrium, where the “optimization strategy” is inherent to the physical processes of Rayleigh–Bénard–Marangoni instabilities and electromagnetic feedback. We call this process “learning” in the broadest sense, deliberately avoiding analogies to specific computational frameworks.

The adaptive evolution of surface morphology under repeated laser irradiation raises a fundamental question: How does matter \textit{learn} to adapt to repeated light exposure? The answer lies in the \textit{emergence of complexity} through hydrodynamically driven transformations \cite{bonse2020maxwell,rudenko2023light}. As the material undergoes repeated interactions with the same light flow, its energy absorption increases alongside its apparent structural complexity, up to a critical threshold. This marks the signature of a learning process, where successive laser irradiation first enhances the material’s sensitivity before triggering a self-organized response that optimizes the source-receiver interface through energy absorption. As the topography evolves, even subtle structural deviations give rise to less efficient configurations, gradually diminishing the material’s ability to harness energy effectively. However, under prolonged exposure, the emerging patterns tailor a state of absorption that is overly optimized, leading to degradation due to pronounced local effects that undermine the overall structure. 

%Learning regime and chaotic regime (Intro)
As surface self-organization progresses and patterns evolve as shown in Fig.~\ref{main-fig:figure1}, complexity and optical response influence each other reciprocally. By associating the dynamics of \( E_{\text{abs}} \) and \( \mathscr{C} \) in Fig.~\ref{fig:figure2}, two successive behavioral regimes are identified. In the \textit{learning regime}, absorbed energy correlates strongly with pattern complexity under fixed laser conditions. Beyond a threshold, a \textit{chaotic regime} arises, where local curvature effects dominate light absorption at the nanoscale, breaking this correlation.

The analysis of Fig.\ref{fig:figure2} indicates that during \\self-organization, the surface acquires new functionalities that improve its sensitivity to light. Furthermore, Fig.\ref{fig:figure3} reveals that while absorption increases for all polarization angles, the material exhibits the highest and lowest sensitivity at \( \alpha=0 \) or \( \alpha=\frac{\pi}{2} \), corresponding to the experimental polarization angles. This response resembles sensitization, a form of non-associative learning in which repeated exposure to a stimulus enhances sensitivity\cite{kandel2000principles,poon2006nonassociative}. Unlike associative learning, where an organism learns to associate two stimuli, non-associative learning involves changes in response to a stimulus over time. While, in living organisms, sensitization arises from neuroplasticity, the brain's ability to reorganize its neural network connections in response to new experiences\cite{grafman2000conceptualizing}. In irradiated materials, this manifests through the formation of patterns that exhibit enhanced sensitivity to the light conditions, particularly the polarization to which they were exposed.

Furthermore, at the microscopic scale, surface structuring begins with localized energy absorption, initiating modifications that seed pattern development. This marks the \emph{response} phase (I), where the surface detects and reacts to laser irradiation by developing roughness. A similar process occurs in plants where low-light environments trigger survival mechanisms. As the interaction continues, both systems enter the \emph{iterative learning} phase (II). The transition from Response to Iterative Learning is driven by the emergence of organized thermoconvective instabilities within the laser-melted layer. The transition occurs when two conditions are simultaneously met: sufficient surface roughness amplitude to create significant near-field electromagnetic inhomogeneities, and thermal conditions that support the development of Rayleigh-Bénard-Marangoni convection within the molten layer lifetime.
Because, for materials, morphological evolution is driven by dissipated energy, surface modulations that locally enhance optical absorption induce stronger thermal gradients and material transport, and therefore grow at higher rates. This nonlinear feedback leads to gradual self-organization of the surface into resonant modes that efficiently couple light into the material. A closely related example is provided by the formation of snow penitentes, where pointed surface structures emerge under solar irradiation due to differential radiative absorption and sublimation, resulting from a similar feedback between surface geometry and energy dissipation \cite{bergeron2006controlled}. In the same way, plants progressively adjust their structure through photomorphogenesis\cite{nemhauser2002photomorphogenesis, schmitt1997photomorphogenic}, resulting in stem elongation and modified leaf orientation to maximize the capture of incoming light \cite{diaz2018mediterranean, wang2022leaf}. With further exposure, they reach the \emph{memory stabilization} phase (III). The transition from Iterative Learning to Memory Stabilization represents kinetic arrest of morphological evolution—the freezing of hydrodynamic pattern dynamics into quasi-stable configurations. The transition occurs when the timescale for pattern reorganization through convection exceeds the melt lifetime, causing the surface morphology to become effectively frozen in an optimized state. In both materials and plants, previous modifications are permanently integrated into the survival strategy, making the process irreversible due to hysteresis. However, persistence in this condition leads to the \emph{destruction} phase (IV). The transition from Memory Stabilization to Destruction reflects the destabilization threshold where accumulated energy deposition exceeds reversible, structure-preserving processes. Enhanced absorption at optimized nanostructures leads to thermal runaway at hotspots, where nanoscale features reach temperatures approaching the boiling point. This triggers material removal through evaporation or explosive boiling, while extreme thermal gradients generate stress fields exceeding material strength. In materials, thermal and mechanical stresses erase established structures, reducing absorption efficiency. While in plants, since light serves as both an energy source and a guide for growth, severe deprivation can be fatal. This succession of phases highlights a learning-like process during self-organization in laser-matter interaction. While materials do not consciously learn like organisms, continuous laser exposure reshapes their structure leading, similar to sensitization in living systems. We use "learn" here to align with the broader definition of learning as adaptation based on experience to improve performance.

%Conclusion

To summarize, we demonstrated that self-organized structures resulting from repeated exposure to laser irradiation are a consequence of a learning process. During this process, matter gradually adapts to laser pulses by rearranging itself to create the most adequate structures to locally maximize absorbed energy. We assimilate this adaptive dynamic process to a form of learning, where the material "learns" to respond to a series of laser pulses by reorganizing its structure, enhancing its sensitivity to laser irradiation. The resulting self-organized patterns represent the most efficient configurations for energy absorption. This is comparable to, e.g., reinforcement learning observed in plants, which adapt their structure through interactions with light in the environment.

% \begin{itemize}
%     \item (Verifier Taylor Shannon entropy figure 2) DONE
% \end{itemize}

\section*{Methods}

%epsilon $\epsilon$ , varepsilon $\varepsilon$, mathcal $\mathscr{H}$, mathscr $\mathscr{H}$
\subsection*{Ultrafast laser setup} Capturing inter-pulse structuring dynamics with nanometric resolution has not been achieved for multi-pulse patterning. Previous synchrotron and XFEL experiments require thin films that are destroyed with each pulse \cite{bonse2024probing}, limiting their applicability for bulk materials, where thermal relaxation and rarefaction waves play a crucial role. Additionally, optical super-resolution techniques lack the spatial resolution required to capture inter-pulse structuring dynamics \cite{aguilar2017situ}. To overcome this challenge, we developed a method reconstructing inter-pulse dynamics by shifting the sample after each pulse sequence. 

The ultrafast laser irradiation setup employed Mach-Zehnder interferometry to combine the effects of polarization mismatch with a tunable inter-pulse delay ($\Delta$t), enabling precise control over surface topography at the nanometer scale \cite{Nakhoul2021Apr}. By disrupting the surface isotropy created by a single polarization state, we generated various self-organization patterns on a nickel monocrystal oriented along the (001) axis. A cross-polarization technique was specifically used, with a polarization angle of $\alpha = \frac{\pi}{2}$ and time delays ranging from 0 to 40 ps, as illustrated in Fig.~\ref{main-fig:figure1}. The pulse duration was fixed at 150 fs, and the laser fluence was carefully controlled by adjusting the number $N$ of double-pulse sequences. Prior to laser irradiation, the nickel surface was mechanically polished to an Ra < 5 nm to ensure that the surface dynamics were dominated by a hydrodynamic process, smoothing the inhomogeneous electromagnetic response.

For each key parameter set (laser fluence $F$ and inter-pulse delay $\Delta t$), the irradiation experiments were repeated at least three times across different locations on the same prepared surface, and additionally verified across multiple independently prepared samples, ensuring the reproducibility of the observed dynamics. This multi-level validation strategy allowed us to distinguish between intrinsic variations inherent to nonlinear dynamic processes and variations arising from sample preparation inconsistencies.

\subsection*{Solving Maxwell's equations in inhomogeneous surface} 

Maxwell's equations \cite{torrance1982dynamical} are formulated as follows:
\begin{align}
    \nabla \times \vec{E} = -\frac{\partial \vec{B}}{\partial t},
    \hspace{0.5cm}
    \nabla \times \vec{H} = \frac{\partial \vec{D}}{\partial t},
\end{align}
where $\vec{E}(x,y,z,t)$ and $\vec{H}(x,y,z,t)$ are the electric and magnetic field vectors $\vec{B}=\mu_{0}\vec{H}$ with $\mu_0$ being the vacuum permeability. The medium response is expressed through the displacement field $\vec{D}$ in the frequency domain $\tilde{D}=\epsilon_0 \epsilon \tilde{E}$ where $\tilde{\hspace{0.2cm}}$ represents the temporal spectrum, $\epsilon_0$ is the vacuum permittivity and $\epsilon$ is the permittivity of the medium, with $\epsilon=1$ for vacuum and $\epsilon=\tilde{n}^{2}=(2.15 + 4.3i)^{2}$ for nickel. 

Since we consider a linearly polarized laser pulse propagating along the z-axis, we express the electric field vector as:
\begin{align}
    \vec{E} = E_{x}^{0}\sin(\alpha)\vec{u_{x}} + E_{y}^{0}\cos(\alpha)\vec{u_{y}},
\end{align}
where $E_{x}^{0}$ and $E_{y}^{0}$ represent the amplitudes of electric field components along $x$ and $y$ axes respectively and $\alpha$ is the polarization angle in $xy$-plane measured from the y-axis toward the x-axis.

Maxwell's equations are solved numerically using the Finite-Difference Time-Domain method, from which the energy dissipation rate is derived as follows\cite{zhang2015coherence,fedorov2024light}:
\begin{align}
     \varepsilon_{abs}(x,y,z) = \frac{1}{2}c\epsilon_0 \frac{4 \pi \Im(\tilde{n})}{\lambda}\Re(\tilde{n})\left|\tilde{E}(x,y,z)\right|^{2},
\end{align}
where $\lambda$ is the wavelength in vacuum and $\tilde{n}(x,y,z)$ the local complex refractive index. Optical indices corresponding to an effective electron temperature of approximately 10000 K are employed, which only marginally modifies the cold optical data under relatively low photoexcitation levels\cite{bevillon2018nonequilibrium, silaeva2021drude}.

\begin{comment}
\subsection*{Maxwell's equations surrogate model} 
Given the large number of artificial surfaces we generated, using a numerical solver to calculate energy absorption $\varepsilon_{abs}(x,y)$ would be too time-consuming. Therefore, a surrogate neural network model becomes a more practical alternative. Since, once trained, it can provide fast predictions for large datasets. Initially, 3D SEM images $u_N(x,y,z)$ are fed into the numerical solver to compute $\varepsilon_{abs}(x,y,z)$, with the third dimension represented by grayscale pixel contrast corresponding to height. Both quantities are integrated along the z-direction to construct an image-to-image regression task. For this task, the U-Net neural network \cite{ronneberger2015unet}, originally designed for semantic segmentation in biomedical image analysis, is selected. Its architecture captures both local and global information efficiently, making it well-suited for tasks where precise localization is essential. The U-Net architecture is ultimately repurposed for regression to predict $\varepsilon_{abs}(x,y)$ knowing $u_N(x,y)$ and the polarization angle $\alpha$.  
\end{comment}

\subsection*{Maxwell's equations surrogate model}
Given the large number of artificial surfaces we generated, using a numerical solver to calculate energy absorption $\varepsilon_{abs}(x,y)$ would be too time-consuming. Therefore, a surrogate neural network model becomes a more practical alternative. Since, once trained, it can provide fast predictions for large datasets. The energy absorption map $\varepsilon_{abs}(x,y)$ is then predicted using a neural network model trained on pattern height $u_N(x,y)$ and the incident polarization angle $\alpha$ \cite{banna2024physics}. \\
We use the four experimental series that display the emergence of self-organized surface patterns (see Fig.~1). For each irradiated surface, we computed the corresponding 2D absorption map $\varepsilon_{abs}(x,y)$ for a given polarization angle \( \alpha=k\pi/6 \) for \( k = 0 \) to \( 5 \). This produced datasets\\ $(u_N(x,y),\, \varepsilon_{abs}(x,y))$.
To increase the amount of training data, each SEM image was augmented using a sliding-window, producing 9 partially overlapping $320 \times 320$ crops per image.\\
Their performance was assessed through a leave-one-series-out cross-validation performed over the four SEM series. Each series in turn plays the role of the test set, while the remaining three are split as follows:
\begin{itemize}
    \item Two series are used for training the models across all hyperparameter configurations.  
    \item The third series is used as a validation set to assess accuracy for each configuration.  
    \item This process is repeated across three folds, each time permuting the role of the validation set.  
\end{itemize}

However, the model is applied not to cropped $320 \times 320$ patches but to the full $800 \times 800$ irradiated surfaces. Because the U-Net is fully convolutional, it can process images of arbitrary size without retraining; the same learned filters are reused at all spatial scales. This approach assumes that the statistical distribution of local features in the training patches is representative of those across the full surface. While this enables efficient large-scale absorption prediction, it may underperform on rare or highly extended structures that were under-represented in the training crops. Nevertheless, the model captures the dominant absorption trends sufficiently well to provide a reliable global estimate of total absorbed energy.

\subsection*{Mutual information}
Mutual information \cite{kraskov2004estimating} measures the linear and non-linear dependency between two random variables, quantifying how much information one provides about the other. It is related to entropy, with higher values indicating greater uncertainty reduction. A value of zero is equivalent to the variables being independent. For two continuous random variables $E_{abs}$ (absorbed energy) and $\mathscr{C}$ (complexity), the mutual information between them is given by \cite{cover1991entropy}:
\begin{align}
MI(E_{\text{abs}}; \mathscr{C}) = \sum_{e,c}  P_{E_{\text{abs}}\mathscr{C}}(e, c) \log \left( \frac{P_{E_{\text{abs}}\mathscr{C}}(e, c)}{P_{E_{\text{abs}}}(e) P_{\mathscr{C}}(c)} \right) 
\end{align}
where $P_{E_{\text{abs}}}(e)$ and  $P_{\mathscr{C}}(c)$ are the marginal probability densities of $E_{\text{abs}}$ and $\mathscr{C}$, respectively, and $P_{E_{\text{abs}}\mathscr{C}}(e, c)$ is the joint probability density of $E_{\text{abs}}$ and $\mathscr{C}$. Note that $MI$ is shown for $N\in [2-50]$, as at least 3 observations are required per dataset to be computed.

\subsection*{Taylor complexity}
Self-organized surface patterns exhibit variations in intensity as well as characteristic local slopes and curvatures that are important to their structure. To better quantify the degree of diversity and structure, we use Taylor complexity, an extension of LMC complexity that incorporates structural information via spatial derivatives.\\
López-Ruiz, Mancini, and Calbet defined a measure of statistical complexity now known as \textit{LMC complexity} based on the insight that complex systems strike a balance between diversity ("information", $\mathscr H$) and order ("disequilibrium", $\mathscr D$). Specifically, given a system with $N$ accessible states, and $\mathbf{p}=\left\{p_{i}\right\}_{i=1\ldots N}$ the set of probabilities associated with each state, one quantifies each of these quantities as
\begin{align}
\mathscr H = -K\sum_{i=1}^{N} p_{i} \log p_{i},\quad \mathscr{D}=\sum_{i=1}^{N}\left(p_{i}-\frac{1}{N}\right)^2,
\end{align}
$\mathscr{H}$ being the Shannon entropy of $\mathbf{p}$ ($K$ typically being set to $\frac{1}{\log N}$ to normalize $\mathscr H$ to $[0,1]$), and $\mathscr{D}$ the quadratic distance of $\mathbf{p}$ to equiprobability. \textit{LMC complexity} is defined as the product of these two quantities:
\begin{equation}
    \label{eq:lmc-complexity}
    C = \mathscr{H} \times \mathscr{D},
\end{equation}
This quantity is zero for perfectly ordered systems (e.g. a crystal) and for perfectly disordered systems (e.g. an ideal gas), and maximal for systems that are somewhere in between. Its intuitive interpretation, ease of computation, and demonstrated versatility across a wide range of applications, including image processing, fundamental physics, and biomedical systems~\cite{lopez1995statistical}\\ The choice of measure of diversity and disequilibrium is not unique, each corresponding to different notions of complexity. Crucially, this measure depends on the choice of state space, which corresponds to the choice of statistical description of the system. In this article, we seek to estimate \textit{dynamical} complexity from single SEM images. To that effect, we assume ergodicity and regard each image as the long-term evolution of a collection of slightly perturbed initial conditions under the same local physical process. Estimating dynamical diversity is generally intractable: we considerably simplify the problem by collapsing field states $u$ onto \textit{sign pattern} states $Pu$ via a projection operator $P$, in the manner of permutation entropy~\cite{bandt02_permut_entrop}, which we extend to the multidimensional case: given a field $u(x,y)$ representing a self-organized pattern, we examine all overlapping $3 \times 3$ patches of the image. This patch size is chosen as the minimal neighborhood that allows meaningful evaluation of local slopes and curvatures, while keeping the number of possible sign patterns manageable. For each patch, we first subtract its mean intensity so that only local deviations remain. Each pixel in the patch is then assigned a binary sign $\{+,-\}$ depending on whether its value is above or below the mean, yielding a \textbf{sign-pattern field} $Pu$ with $2^9$ possible configurations per patch. In order to define \textit{shape} statistics rigorously, we aggregate the LMC complexities of the sign patterns of the field and its derivatives in the same way that they contribute to the value (the shape) of the field at the inter-pixel distance, i.e., using the appropriate Taylor expansion coefficients. We call this measure \textit{Taylor-LMC-$n$ complexity} of $u:\mathbb R^p\rightarrow \mathbb R$ (or simply \textit{complexity} when there is no risk of ambiguity) : 
\begin{align}
\label{eq:taylor-statistics-multidim}
T^n C(Pu) = \sum\limits_{|\alpha|=0}^{n} \frac{C(Pu^{(\alpha)})}{\alpha!},
\end{align}
where $\alpha$ is a multi-index $\alpha =(\alpha _{1},\alpha _{2},\ldots ,\alpha _{n})$ of length $n$, with $|\alpha |\ =\ \sum _{k=1}^{n}\alpha _{k}$, the factorial $\alpha !$ denoting $\alpha _{1}!\cdot \alpha _{2}!\cdots \alpha _{n}!$ and $u^{(\alpha)}:=\frac{\partial^{\alpha_1+\cdots \alpha_n}}{\partial x_1^{\alpha_1}\cdots \partial x_n^{\alpha_n}}u$.\\
In this work, we take $n=2$ and construct sign-pattern fields $Pu_x$ and $Pu_y$ from first-order derivatives, as well as $Pu_{xx}$, $Pu_{xy}$, and $Pu_{yy}$ from the second-order derivatives. The Taylor complexity $\mathscr{C}$  is then given by
\begin{align}
\mathscr{C} &=T^{2}C= C(Pu) + C(Pu_x) + C(Pu_y) \nonumber\\
&\quad + \frac{1}{2}C(Pu_{xx}) + C(Pu_{xy}) + \frac{1}{2}C(Pu_{yy}),
\label{eq:taylor_complexity}
\end{align}
where $C(\cdot)$ denotes the LMC complexity computed over the distribution of all $3\times 3$ patches for each field or derivative.\\
See the supplementary material for a detailed presentation.

\subsection*{Creating artificial surfaces}
In order to measure the impact of modifying a pattern that maximizes absorbed energy, we introduce two modification approaches. In the first approach, a patch from the same series, derived from the learning regime, replaces part of the experimental surface. This creates artificial surfaces that, though visually similar to experimental ones, do not follow the physics of self-organization and complexity. In the second approach, a non-linear dynamics model, based on the Swift-Hohenberg equation derived for Rayleigh-Bénard convection \cite{swift1977hydrodynamic}, evolves a patch of the experimental surface. This approach generates artificial surfaces that are physically plausible and consistent with the Swift-Hohenberg equation, which is widely used for modeling pattern formation. We compare the absorbed energy $E_{abs}^{N+1}=f(N,\alpha)$ variation between the original experimental surfaces and their modified counterparts for series 1 (N=31 and $\alpha=\pi/2$), series 2 (N=20 and $\alpha=\pi/2$), series 3 (N=49 and $\alpha=0$) and series 4 (N=25 and $\alpha=0$). In total, 8000 modified surfaces are included, with 4000 surfaces per modification method and 1000 surfaces generated for each method and each series.

\subsection*{Modelling thermoconvective instability}
Patterns formed through a self-organization process can be modeled using the adimensional Swift-Hohenberg equation \cite{brandao23_learn_compl_to_guide_light}, introduced in the context of Rayleigh-Bénard convection \cite{swift1977hydrodynamic}. This partial differential equation describes how systems, far from equilibrium, self-organize into structured patterns under Rayleigh-Bénard convection. The solutions exhibit Turing-like patterns remarkably similar to those observed in laser-irradiated surfaces \cite{brandao23_learn_compl_to_guide_light}. For a field $u(\mathbf{x}, t)$ we have
\begin{align}
\label{eq:sh-eq}
\frac{\partial u}{\partial t} = ru - \left(\mathbf{q}_{0}^2 + \nabla^2\right)^{2} u + \gamma u^2 - u^3,
\end{align}
which we henceforth denote $SH(r,\gamma, \mathbf{q}_0)$, where $r$ is the bifurcation parameter, $\gamma$ breaks the symmetry of the solutions with respect to sign inversion, and $\mathbf{q}_{0}$ is the typical spatial wavenumber that controls the typical scale of the features in the solution.

\section*{Acknowledgements}
This work has been partly funded by a public grant from the French National Research Agency (ANR) under (i) the Investments for the Future Program (PIA), which has the reference EUR MANUTECH SLEIGHT-ANR-17-EURE-0026 and (ii) the MELISSA project with reference ANR-24-CE23-7140-01.

\bibliographystyle{abbrv}
\bibliography{biblio}

@article{hansen2015continuum,
  title={Continuum nanofluidics},
  author={Hansen, Jesper S and Dyre, Jeppe C and Daivis, Peter and Todd, Billy D and Bruus, Henrik},
  journal={Langmuir},
  volume={31},
  number={49},
  pages={13275--13289},
  year={2015},
  publisher={ACS Publications}
}

@article{trice2008novel,
  title={Novel self-organization mechanism in ultrathin liquid films: theory and experiment},
  author={Trice, Justin and Favazza, Christopher and Thomas, Dennis and Garcia, Hernando and Kalyanaraman, <? format?> Ramki and Sureshkumar, Radhakrishna},
  journal={Physical review letters},
  volume={101},
  number={1},
  pages={017802},
  year={2008},
  publisher={APS}
}

@article{bonse2020laser,
  title={Laser-induced periodic surface structures (LIPSS)},
  author={Bonse, J{\"o}rn and Kirner, Sabrina V and Kr{\"u}ger, J{\"o}rg},
  journal={Handbook of laser micro-and nano-engineering},
  pages={1--59},
  year={2020},
  publisher={Springer}
}

@article{her1998microstructuring,
  title={Microstructuring of silicon with femtosecond laser pulses},
  author={Her, Tsing-Hua and Finlay, Richard J and Wu, Claudia and Deliwala, Shrenik and Mazur, Eric},
  journal={Applied Physics Letters},
  volume={73},
  number={12},
  pages={1673--1675},
  year={1998},
  publisher={American Institute of Physics}
}

@article{gattass2008femtosecond,
  title={Femtosecond laser micromachining in transparent materials},
  author={Gattass, Rafael R and Mazur, Eric},
  journal={Nature photonics},
  volume={2},
  number={4},
  pages={219--225},
  year={2008},
  publisher={Nature Publishing Group UK London}
}

@article{zhang2015coherence,
  title={Coherence in ultrafast laser-induced periodic surface structures},
  author={Zhang, Hao and Colombier, Jean-Philippe and Li, Chen and Faure, Nicolas and Cheng, Guanghua and Stoian, Razvan},
  journal={Physical Review B},
  volume={92},
  number={17},
  pages={174109},
  year={2015},
  publisher={APS}
}

@article{ilday2017rich,
  title={Rich complex behaviour of self-assembled nanoparticles far from equilibrium},
  author={Ilday, Serim and Makey, Ghaith and Akguc, Gursoy B and Yavuz, {\"O}zg{\"u}n and Tokel, Onur and Pavlov, Ihor and G{\"u}lseren, Oguz and Ilday, F {\"O}mer},
  journal={Nature communications},
  volume={8},
  number={1},
  pages={14942},
  year={2017},
  publisher={Nature Publishing Group UK London}
}

@article{makey2020universality,
  title={Universality of dissipative self-assembly from quantum dots to human cells},
  author={Makey, Ghaith and Galioglu, Sezin and Ghaffari, Roujin and Engin, E Doruk and Y{\i}ld{\i}r{\i}m, G{\"o}khan and Yavuz, {\"O}zg{\"u}n and Bekta{\c{s}}, Onurcan and Nizam, {\"U} Seleme and Akbulut, {\"O}zge and {\c{S}}ahin, {\"O}zg{\"u}r and others},
  journal={Nature Physics},
  volume={16},
  number={7},
  pages={795--801},
  year={2020},
  publisher={Nature Publishing Group UK London}
}

@article{bonse2024probing,
  title={Probing Laser-Driven Structure Formation at Extreme Scales in Space and Time},
  author={Bonse, J{\"o}rn and Sokolowski-Tinten, Klaus},
  journal={Laser \& Photonics Reviews},
  volume={18},
  number={5},
  pages={2300912},
  year={2024},
  publisher={Wiley Online Library}
}

@article{aguilar2017situ,
  title={In-situ high-resolution visualization of laser-induced periodic nanostructures driven by optical feedback},
  author={Aguilar, Alberto and Mauclair, Cyril and Faure, Nicolas and Colombier, Jean-Philippe and Stoian, Razvan},
  journal={Scientific Reports},
  volume={7},
  number={1},
  pages={16509},
  year={2017},
  publisher={Nature Publishing Group UK London}
}

@article{bonse2020maxwell,
  title={Maxwell meets Marangoni—a review of theories on laser-induced periodic surface structures},
  author={Bonse, J{\"o}rn and Gr{\"a}f, Stephan},
  journal={Laser \& Photonics Reviews},
  volume={14},
  number={10},
  pages={2000215},
  year={2020},
  publisher={Wiley Online Library}
}

@article{fedorov2024light,
  title={Light-matter interaction at rough surfaces: A morphological perspective on laser-induced periodic surface structures},
  author={Fedorov, Vladimir Yu and Colombier, Jean-Philippe},
  journal={Physical Review B},
  volume={110},
  number={11},
  pages={115438},
  year={2024},
  publisher={APS}
}

@article{lu2018influence,
  title={Influence of surface roughness on strong light-matter interaction of a quantum emitter-metallic nanoparticle system},
  author={Lu, Yu-Wei and Li, Ling-Yan and Liu, Jing-Feng},
  journal={Scientific Reports},
  volume={8},
  number={1},
  pages={7115},
  year={2018},
  publisher={Nature Publishing Group UK London}
}

@article{diaz2018mediterranean,
  title={How do Mediterranean shrub species cope with shade? Ecophysiological response to different light intensities},
  author={D{\'\i}az-Barradas, Mari Cruz and Zunzunegui, Mar{\'\i}a and Alvarez-Cansino, Leonor and Esquivias, Mari Paz and Valera, J and Rodr{\'\i}guez, Herminia},
  journal={Plant Biology},
  volume={20},
  number={2},
  pages={296--306},
  year={2018},
  publisher={Wiley Online Library}
}

@article{wang2022leaf,
  title={Leaf morphological traits as adaptations to multiple climate gradients},
  author={Wang, Han and Wang, Runxi and Harrison, Sandy P and Prentice, Iain Colin},
  journal={Journal of Ecology},
  volume={110},
  number={6},
  pages={1344--1355},
  year={2022},
  publisher={Wiley Online Library}
}

@book{lambers2008plant,
  title={Plant physiological ecology},
  author={Lambers, Hans and Chapin III, F Stuart and Pons, Thijs L},
  year={2008},
  publisher={Springer Science \& Business Media}
}

@article{middleton1994understanding,
  title={Understanding photosynthesis, pigment and growth responses induced by UV-B and UV-A irradiances},
  author={Middleton, Elizabeth M and Teramura, Alan H},
  journal={Photochemistry and photobiology},
  volume={60},
  number={1},
  pages={38--45},
  year={1994},
  publisher={Wiley Online Library}
}

@article{teramura1994effects,
  title={Effects of UV-B radiation on photosynthesis and growth of terrestrial plants},
  author={Teramura, Alan H and Sullivan, Joe H},
  journal={Photosynthesis research},
  volume={39},
  pages={463--473},
  year={1994},
  publisher={Springer}
}

@article{chaves2009photosynthesis,
  title={Photosynthesis under drought and salt stress: regulation mechanisms from whole plant to cell},
  author={Chaves, MM and Flexas, J and Pinheiro, Carla},
  journal={Annals of botany},
  volume={103},
  number={4},
  pages={551--560},
  year={2009},
  publisher={Oxford University Press}
}

@misc{maier2007plasmonics,
  title={Plasmonics: Fundamentals and Applications},
  author={Maier, SA},
  year={2007},
  publisher={Springer}
}

@article{ferry2011modeling,
  title={Modeling light trapping in nanostructured solar cells},
  author={Ferry, Vivian E and Polman, Albert and Atwater, Harry A},
  journal={ACS nano},
  volume={5},
  number={12},
  pages={10055--10064},
  year={2011},
  publisher={ACS Publications}
}

@book{novotny2012principles,
  title={Principles of nano-optics},
  author={Novotny, Lukas and Hecht, Bert},
  year={2012},
  publisher={Cambridge university press}
}

@article{penuelas2003bvocs,
  title={BVOCs: plant defense against climate warming?},
  author={Pe{\~n}uelas, Josep and Llusi{\`a}, Joan},
  journal={Trends in plant science},
  volume={8},
  number={3},
  pages={105--109},
  year={2003},
  publisher={Elsevier}
}

@book{sugioka2013ultrafast,
  title={Ultrafast laser processing: from micro-to nanoscale},
  author={Sugioka, Koji and Cheng, Ya},
  year={2013},
  publisher={CRC Press}
}

@article{stoian2020advances,
  title={Advances in ultrafast laser structuring of materials at the nanoscale},
  author={Stoian, Razvan and Colombier, Jean-Philippe},
  journal={Nanophotonics},
  volume={9},
  number={16},
  pages={4665--4688},
  year={2020},
  publisher={De Gruyter}
}

@article{zorba2008biomimetic,
  title={Biomimetic artificial surfaces quantitatively reproduce the water repellency of a lotus leaf},
  author={Zorba, Vassilia and Stratakis, Emmanuel and Barberoglou, Marios and Spanakis, Emmanuel and Tzanetakis, Panagiotis and Anastasiadis, Spiros H and Fotakis, Costas},
  journal={Advanced materials},
  volume={20},
  number={21},
  pages={4049--4054},
  year={2008},
  publisher={Wiley Online Library}
}

@article{shimotsuma2003self,
  title={Self-organized nanogratings in glass irradiated by ultrashort light pulses},
  author={Shimotsuma, Yasuhiko and Kazansky, Peter G and Qiu, Jiarong and Hirao, Kazuoki},
  journal={Physical review letters},
  volume={91},
  number={24},
  pages={247405},
  year={2003},
  publisher={APS}
}

@article{yao2023materials,
  title={Materials roadmap for inscription of nanogratings inside transparent dielectrics using ultrafast lasers},
  author={Yao, Heng and Xie, Qiong and Cavillon, Maxime and Dai, Ye and Lancry, Matthieu},
  journal={Progress in Materials Science},
  pages={101226},
  year={2023},
  publisher={Elsevier}
}

@article{perrakis2024impact,
  title={Impact of Hybrid Electromagnetic Surface Modes on the Formation of Low Spatial Frequency LIPSS: A Universal Approach},
  author={Perrakis, George and Tsilipakos, Odysseas and Tsibidis, George D and Stratakis, Emmanuel},
  journal={Laser \& Photonics Reviews},
  pages={2301090},
  year={2024},
  publisher={Wiley Online Library}
}

@book{wiener2019cybernetics,
  title={Cybernetics or Control and Communication in the Animal and the Machine},
  author={Wiener, Norbert},
  year={2019},
  publisher={MIT press}
}

@article{rudenko2020high,
  title={High-frequency periodic patterns driven by non-radiative fields coupled with {M}arangoni convection instabilities on laser-excited metal surfaces},
  author={Rudenko, Anton and Abou-Saleh, Anthony and Pigeon, Florent and Mauclair, Cyril and Garrelie, Florence and Stoian, Razvan and Colombier, Jean-Philippe},
  journal={Acta Materialia},
  volume={194},
  pages={93--105},
  year={2020},
  publisher={Elsevier}
}

@article{baev2015metaphotonics,
  title={Metaphotonics: {A}n emerging field with opportunities and challenges},
  author={Baev, Alexander and Prasad, Paras N and {\AA}gren, Hans and Samo{\'c}, Marek and Wegener, Martin},
  journal={Physics Reports},
  volume={594},
  pages={1--60},
  year={2015},
  publisher={Elsevier}
}

@article{Stratakis2020Jul,
	author = {Stratakis, E. and Bonse, J. and Heitz, J. and Siegel, J. and Tsibidis, G. D. and Skoulas, E. and Papadopoulos, A. and Mimidis, A. and Joel, A.-C. and Comanns, P. and Kr{\ifmmode\ddot{u}\else\"{u}\fi}ger, J. and Florian, C. and Fuentes-Edfuf, Y. and Solis, J. and Baumgartner, W.},
	title = {{Laser engineering of biomimetic surfaces}},
	journal = {Materials Science and Engineering: R: Reports},
	volume = {141},
	pages = {100562},
	year = {2020},
	month = jul,
	issn = {0927-796X},
	publisher = {Elsevier},
	doi = {10.1016/j.mser.2020.100562}
}

@article{zhang2020laser,
  title={Laser-induced periodic surface structures: Arbitrary angles of incidence and polarization states},
  author={Zhang, Hao and Colombier, Jean-Philippe and Witte, Stefan},
  journal={Physical Review B},
  volume={101},
  number={24},
  pages={245430},
  year={2020},
  publisher={APS}
}

@article{rudenko2019self,
  title={Self-organization of surfaces on the nanoscale by topography-mediated selection of quasi-cylindrical and plasmonic waves},
  author={Rudenko, Anton and Mauclair, Cyril and Garrelie, Florence and Stoian, Razvan and Colombier, Jean-Philippe},
  journal={Nanophotonics},
  volume={8},
  number={3},
  pages={459--465},
  year={2019},
  publisher={De Gruyter}
}

@article{tsibidis2015ripples,
  title={From ripples to spikes: A hydrodynamical mechanism to interpret femtosecond laser-induced self-assembled structures},
  author={Tsibidis, George D and Fotakis, Costas and Stratakis, Emmanuel},
  journal={Physical Review B},
  volume={92},
  number={4},
  pages={041405},
  year={2015},
  publisher={APS}
}

@incollection{rudenko2023light,
  title={How light drives material periodic patterns down to the nanoscale},
  author={Rudenko, Anton and Colombier, Jean-Philippe},
  booktitle={Ultrafast Laser Nanostructuring: The Pursuit of Extreme Scales},
  pages={209--255},
  year={2023},
  publisher={Springer}
}

@article{prigogine1963introduction,
  title={Introduction to thermodynamics of irreversible processes},
  author={Prigogine, Ilya and Van Rysselberghe, Pierre},
  journal={Journal of The Electrochemical Society},
  volume={110},
  number={4},
  pages={97C},
  year={1963},
  publisher={IOP Publishing}
}

@article{swift1977hydrodynamic,
  title={Hydrodynamic fluctuations at the convective instability},
  author={Swift, Ju and Hohenberg, Pierre C},
  journal={Physical Review A},
  volume={15},
  number={1},
  pages={319},
  year={1977},
  publisher={APS}
}

@phdthesis{PhD-Brandao-2024,
  author =	 {Eduardo Brandao},
  title =	 {Complexity Methods in Physics-Guided Machine
                  Learning},
  school =	 {Jean Monnet University, France},
  year =	 2024
}

@misc{ronneberger2015unet,
      title={U-Net: Convolutional Networks for Biomedical Image Segmentation}, 
      author={Olaf Ronneberger and Philipp Fischer and Thomas Brox},
      year={2015},
      eprint={1505.04597},
      archivePrefix={arXiv},
      primaryClass={cs.CV}
}

@article{valtr95_probab_that_random_point_are_convex_posit,
  author =	 {Valtr, P.},
  title =	 {Probability That Random Points Are in Convex Position},
  journal =	 {Discrete \& Computational Geometry},
  volume =	 13,
  number =	 3,
  pages =	 {637--643},
  year =	 1995,
  doi =		 {10.1007/BF02574070},
  url =		 {https://doi.org/10.1007/BF02574070},
  issn =	 {1432-0444},
}

@Article{brandao22_learn_pde_to_model_self_organ_matter,
  title =	 {Learning Pde To Model Self-Organization of Matter},
  ARTICLE-NUMBER =1096,
  AUTHOR =	 {Brandao, Eduardo and Colombier, Jean-Philippe and Duffner,
                  Stefan and Emonet, R{\'e}mi and Garrelie, Florence and Habrard,
                  Amaury and Jacquenet, Fran{\c{c}}ois and Nakhoul, Anthony and
                  Sebban, Marc},
  DOI =		 {10.3390/e24081096},
  ISSN =	 {1099-4300},
  JOURNAL =	 {Entropy},
  NUMBER =	 8,
  PubMedID =	 36010759,
  URL =		 {https://www.mdpi.com/1099-4300/24/8/1096},
  VOLUME =	 24,
  YEAR =	 2022,
}

@article{brandao23_learn_compl_to_guide_light,
  author =	 {Brandao, Eduardo and Nakhoul, Anthony and Duffner, Stefan and
                  Emonet, R. and Garrelie, Florence and Habrard, Amaury and
                  Jacquenet, Fran{\c{c}}ois and Pigeon, Florent and Sebban, Marc
                  and Colombier, Jean-Philippe},
  title =	 {Learning Complexity To Guide Light-Induced Self-Organized
                  Nanopatterns},
  journal =	 {Physical Review Letters},
  volume =	 130,
  number =	 22,
  year =	 2023,
  url =		 {http://dx.doi.org/10.1103/PhysRevLett.130.226201},
  DOI =		 {10.1103/physrevlett.130.226201},
  ISSN =	 {1079-7114},
  month =	 may,
  publisher =	 {American Physical Society (APS)},
}

@article{Nakhoul2021Apr,
	author = {Nakhoul, Anthony and Maurice, Claire and Agoyan, Marion and Rudenko, Anton and Garrelie, Florence and Pigeon, Florent and Colombier, Jean-Philippe},
	title = {{Self-Organization Regimes Induced by Ultrafast Laser on Surfaces in the Tens of Nanometer Scales}},
	journal = {Nanomaterials},
	volume = {11},
	number = {4},
	pages = {1020},
	year = {2021},
	month = apr,
	issn = {2079-4991},
	publisher = {Multidisciplinary Digital Publishing Institute},
	doi = {10.3390/nano11041020}
}

@article{lopez1995statistical,
  title={A statistical measure of complexity},
  author={Lopez-Ruiz, Ricardo and Mancini, Hector L and Calbet, Xavier},
  journal={Physics letters A},
  volume={209},
  number={5-6},
  pages={321--326},
  year={1995},
  publisher={Elsevier}
}

@article{parzen1962estimation,
  title={On estimation of a probability density function and mode},
  author={Parzen, Emanuel},
  journal={The annals of mathematical statistics},
  volume={33},
  number={3},
  pages={1065--1076},
  year={1962},
  publisher={JSTOR}
}

@article{Liu1982May,
	author = {Liu, J. M.},
	title = {{Simple technique for measurements of pulsed Gaussian-beam spot sizes}},
	journal = {Opt. Lett.},
	volume = {7},
	number = {5},
	pages = {196--198},
	year = {1982},
	month = {May},
	issn = {1539-4794},
	publisher = {Optica Publishing Group},
	doi = {10.1364/OL.7.000196}
}

@article{nemhauser2002photomorphogenesis,
  title={Photomorphogenesis},
  author={Nemhauser, Jennifer and Chory, Joanne},
  journal={The Arabidopsis Book/American Society of Plant Biologists},
  volume={1},
  year={2002},
  publisher={American Society of Plant Biologists}
}

@article{schmitt1997photomorphogenic,
  title={Is photomorphogenic shade avoidance adaptive? Perspectives from population biology},
  author={Schmitt, J},
  journal={Plant, Cell \& Environment},
  volume={20},
  number={6},
  pages={826--830},
  year={1997},
  publisher={Wiley Online Library}
}

@article{grafman2000conceptualizing,
  title={Conceptualizing functional neuroplasticity},
  author={Grafman, Jordan},
  journal={Journal of communication disorders},
  volume={33},
  number={4},
  pages={345--356},
  year={2000},
  publisher={Elsevier}
}

@book{kandel2000principles,
  title={Principles of neural science},
  author={Kandel, Eric R and Schwartz, James H and Jessell, Thomas M and Siegelbaum, Steven and Hudspeth, A James and Mack, Sarah and others},
  volume={4},
  year={2000},
  publisher={McGraw-hill New York}
}

@article{poon2006nonassociative,
  title={Nonassociative learning as gated neural integrator and differentiator in stimulus-response pathways},
  author={Poon, Chi-Sang and Young, Daniel L},
  journal={Behavioral and Brain Functions},
  volume={2},
  pages={1--21},
  year={2006},
  publisher={Springer}
}

@article{kraskov2004estimating,
  title={Estimating mutual information},
  author={Kraskov, Alexander and St{\"o}gbauer, Harald and Grassberger, Peter},
  journal={Physical Review E-Statistical, Nonlinear, and Soft Matter Physics},
  volume={69},
  number={6},
  pages={066138},
  year={2004},
  publisher={APS}
}

@article{cover1991entropy,
  title={Entropy, relative entropy and mutual information},
  author={Cover, Thomas M and Thomas, Joy A and others},
  journal={Elements of information theory},
  volume={2},
  number={1},
  pages={12--13},
  year={1991}
}

@misc{torrance1982dynamical,
  title={A Dynamical Theory of the Electromagnetic Field},
  author={Torrance, TF and Maxwell, James Clerk},
  year={1982},
  publisher={Edinburgh: Scottish Academic Press}
}

@article{shannon2001mathematical,
  title={A mathematical theory of communication},
  author={Shannon, Claude Elwood},
  journal={ACM SIGMOBILE mobile computing and communications review},
  volume={5},
  number={1},
  pages={3--55},
  year={2001},
  publisher={ACM New York, NY, USA}
}

@article{bandt02_permut_entrop,
	title        = {Permutation Entropy: a Natural Complexity Measure for Time Series},
	author       = {Bandt, Christoph and Pompe, Bernd},
	year         = 2002,
	journal      = {Physical review letters},
	publisher    = {APS},
	volume       = 88,
	number       = 17,
	pages        = 174102
}

@book{khinchin2013mathematical,
  title={Mathematical foundations of information theory},
  author={Khinchin, A Ya},
  year={2013},
  publisher={Courier Corporation}
}

@misc{lopezruiz2010statisticalmeasurecomplexity,
      title={A Statistical Measure of Complexity}, 
      author={Ricardo Lopez-Ruiz and Hector Mancini and Xavier Calbet},
      year={2010},
      eprint={1009.1498},
      archivePrefix={arXiv},
      primaryClass={nlin.AO},
      url={https://arxiv.org/abs/1009.1498}, 
}

@inproceedings{banna2024physics,
  title={Physics-informed Machine Learning for Better Understanding Laser-Matter Interaction},
  author={Banna, Fayad Ali and Colombier, Jean-Philippe and Emonet, R{\'e}mi and Sebban, Marc},
  booktitle={2024 IEEE 36th International Conference on Tools with Artificial Intelligence (ICTAI)},
  pages={199--205},
  year={2024},
  organization={IEEE}
}

@article{bevillon2018nonequilibrium,
  title={Nonequilibrium optical properties of transition metals upon ultrafast electron heating},
  author={B{\'e}villon, Emile and Stoian, R and Colombier, Jean-Philippe},
  journal={Journal of Physics: Condensed Matter},
  volume={30},
  number={38},
  pages={385401},
  year={2018},
  publisher={IOP Publishing}
}

@article{silaeva2021drude,
  title={Drude-lorentz model for optical properties of photoexcited transition metals under electron-phonon nonequilibrium},
  author={Silaeva, Elena and Saddier, Louis and Colombier, Jean-Philippe},
  journal={Applied Sciences},
  volume={11},
  number={21},
  pages={9902},
  year={2021},
  publisher={MDPI}
}

@article{bergeron2006controlled,
  title={Controlled irradiative formation of penitentes},
  author={Bergeron, Vance and Berger, Charles and Betterton, MD},
  journal={Physical review letters},
  volume={96},
  number={9},
  pages={098502},
  year={2006},
  publisher={APS}
}

\end{document}

\maketitle              

\section{Experimental process \& nanostructuring}

\begin{figure}[h!]
    \centering
    \includegraphics[width=8cm]{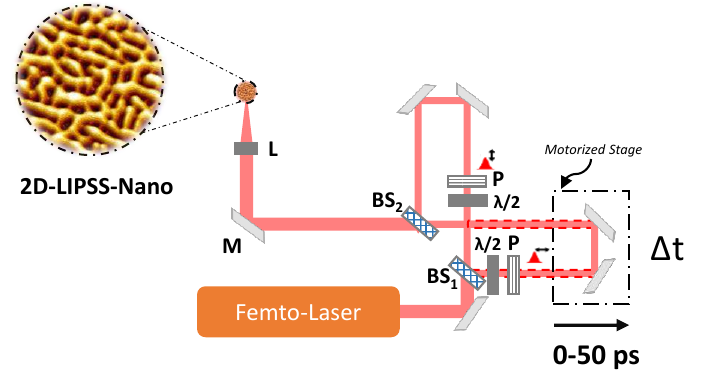}
    \captionsetup{justification=justified, singlelinecheck=false} 
    \caption{\textbf{Schematic illustration of a femtosecond laser double-pulse setup with temporal control.} BS1 and BS2 refer to the beam splitters, $\lambda/2$ represents the half-wave plate, P denotes the polarizer, M the mirror, and L the lens, which has a focal distance of 25 cm from the sample.}
    \label{supp-fig:fgr1}
\end{figure}

A Coherent Legend Elite Series Ti:Sapphire laser was used for the experiments. This system provides linearly polarized pulses with a minimum pulse duration of approximately 40 fs. The ultrafast amplifier has a repetition rate of 1 kHz, an output power of 3 W, and a central wavelength of 800 nm. The ultrashort pulses are generated and amplified by the pump laser and are ejected using a pulse picker. To induce self-organization on the surface, cross-polarized irradiation was employed, achieved by modifying a Mach–Zehnder interferometer, which enables isotropic energy deposition.

Fig.~\ref{supp-fig:fgr1} presents a schematic illustration of the experimental setup for cross-polarized laser irradiation. Initially, the incoming laser beam is split by a non-polarizing beam splitter (BS1) with a 50/50 ratio. The two beams then follow separate optical paths, or arms, before recombining at a second beam splitter (BS2). The optical path length of arm one is controlled by a motorized stage, which enables temporal control between the two beams in the picosecond regime. The energy and polarization of both beams can be independently controlled using half-wave plates ($\lambda/2$) and polarizers. When the path lengths of the two arms are equal, temporal overlap ($\Delta t = 0$) occurs. This overlap is verified using an autocorrelator when the polarization angle is set to $0^\circ$.

Each optical arm is equipped with a polarizer (P) to control the polarization angle, ensuring the desired cross-polarization. The cross-polarization condition is confirmed using a polarizer, allowing precise and quick adjustments. The light from the interferometer is then recombined, collimated using a 25 cm focal length lens, and directed to the sample at normal incidence. The laser spot size, with a Gaussian profile (at $1/e^2$), has a diameter of approximately 60~\textmu m, determined by the beam waist ($\omega_0$). The spot size is carefully measured before irradiation using the D\textsuperscript{2} method~\cite{Liu1982May}.

\begin{figure}[h!]
    \centering
    \includegraphics[width=8cm]{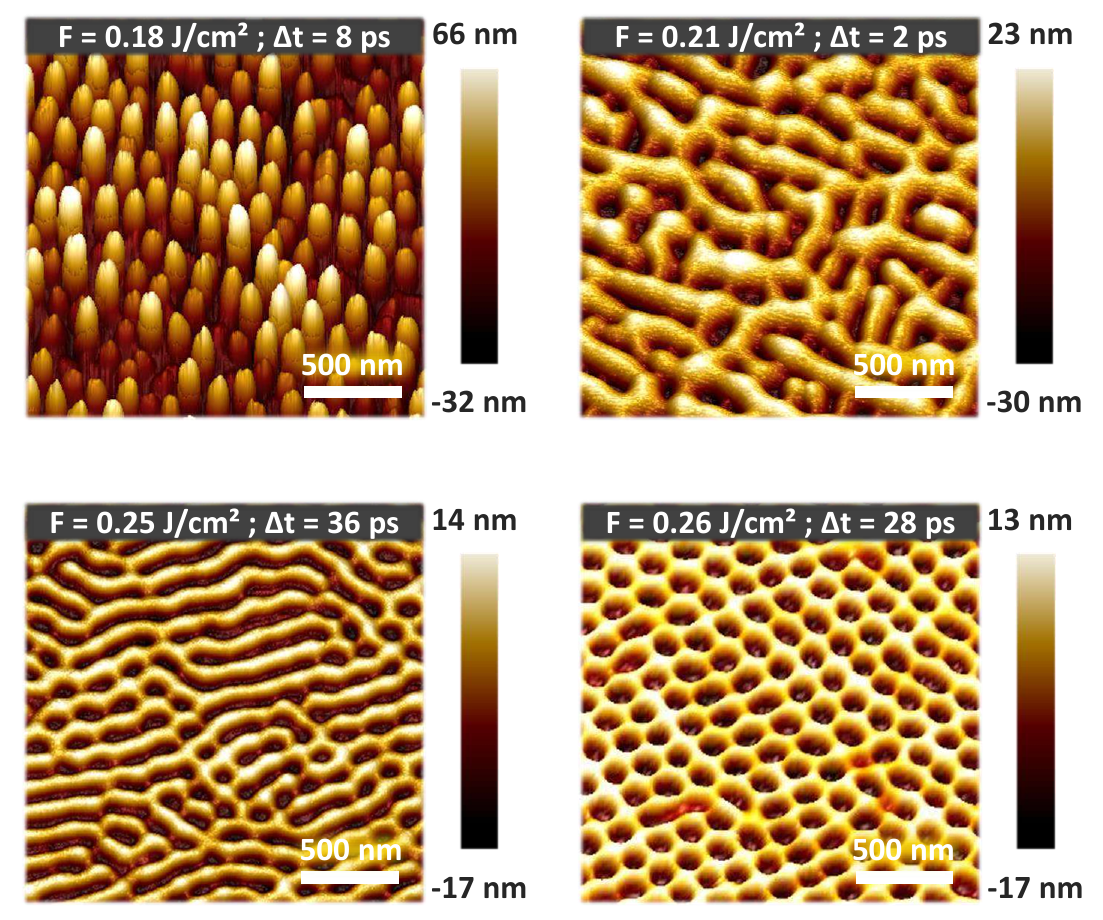}
    \captionsetup{justification=justified, singlelinecheck=false} 
    \caption{\textbf{Atomic Force Microscopy (AFM) images of laser-induced nanostructures formed on Ni(001)}. The structures form at different laser parameters with $N = 25$ double pulse sequences. The observed nanostructures include nanopeaks, nanowebs, nanolabyrinths, and nanocavities. These structures form progressively at different laser doses: (0.18 J/cm\textsuperscript{2}, 8 ps), (0.21 J/cm\textsuperscript{2}, 2 ps), (0.25 J/cm\textsuperscript{2}, 36 ps), and (0.26 J/cm\textsuperscript{2}, 28 ps). The height of these nanostructures varies from tens of nanometers up to approximately 100 nm.}
    \label{fgr2}
\end{figure}

%Fig.~\ref{fgr2} presents AFM images of laser-induced nanostructures on Ni(001) at different laser parameters, providing a height characterization complementary to the SEM observations in the main text. AFM measurements confirm the formation of different types of nanostructures—nanopeaks, nanowebs, nanolabyrinths, and nanocavities—which correlate with variations in fluence $F$ and time delay $\Delta t$. The structures evolve progressively with increasing dose, highlighting the interplay between laser-induced convection dynamics and self-organization. Notably, the height of these nanostructures ranges from ten to about one hundred nanometers, with nanopeaks exhibiting the highest aspect ratio. These findings support the feedback-driven morphological transitions described in the main text.

Fig.~\ref{fgr2} presents AFM images of laser-induced nanostructures on Ni(001) at different laser parameters, providing a height characterization complementary to the SEM observations in the main text. AFM measurements confirm the formation of different types of nanostructures, including nanopeaks, nanowebs, nanolabyrinths, and nanocavities, which correlate with variations in fluence $F$ and time delay $\Delta t$. The structures evolve progressively with increasing dose, highlighting the interplay between laser-induced convection dynamics and self-organization. Notably, the height of these nanostructures ranges from ten to about one hundred nanometers, with nanopeaks exhibiting the highest aspect ratio. These findings support the feedback-driven morphological transitions described in the main text.

\section{Definition of complexity}
In this section we present \textit{Taylor Complexity}, a measure of dynamical complexity which incorporates shape diversity information via the \textit{derivative pattern distribution}, originally proposed by~\cite{PhD-Brandao-2024} in the context of machine learning for self-organization. The measure extends permutation entropy \cite{bandt02_permut_entrop} and is inspired by statistical complexity introduced in \cite{lopez1995statistical}. It is straightforward to compute, interpret, and potentially applicable to broader domains.

We begin by noting that complexity is easy to identify but difficult to define. A number of complexity measures have been proposed in the literature, ranging from computational complexity, such as Kolmogorov complexity or Lempel-Ziv entropy; to information-theoretical measures such as Shannon entropy; measures of dynamical complexity in the sense of chaos, such as Lyapunov exponents or Kolmogorov-Sinai entropy, or fractal dimension; or a number of ad-hoc measures tailored to specific problems or fields.

The difficulty in defining a measure of complexity that is suitable to a given problem arises from the need to clarify the three different points: (i) \textit{what exactly are we measuring the complexity of?} (ii) \textit{what exactly do we mean by complexity?} (iii) \textit{what measure are we going to use?} We shall examine each of these points in turn.

\paragraph{{(i) Measuring complexity of self-organization dynamics from a single SEM image}}The main thrust of this report is that matter self-organizes to maximize energy absorption, and that this can be seen as a learning process, in the sense of the sense of~\cite{wiener2019cybernetics}. For that reason, we seek a measure of \textit{dynamical} complexity that is applicable to the self-organization process. Moreover, this measure should be estimated via \textit{a single snapshot} of that dynamics given as an SEM image (cf. Fig.~1).

% Our focus is on measures of complexity relevant to self-organization and pattern formation, particularly in relation to ultrafast laser-induced nanopatterns. 

SEM images analyzed in this study are characterized by \textit{patterns}, i.e. structures that present regularity but that are diverse to some degree. Importantly, each structure is formed under quasi-uniform initial conditions via the same hydrodynamic self-organization process. Concretely, we consider each SEM image, represented by a scalar field $u(\mathbf{x}, t)$, as a collection of patches resulting from independent, long-term evolutions of slightly perturbed initial conditions and averaged boundary conditions, all governed by the same local process $\frac{\partial u}{\partial t} = F(u)$, where $F$ is a functional of the field $u$, a number of its spatial derivatives, and potentially terms depending on $x$ and $t$. In the absence of long-range interactions, the correlations between different structures arise from a composition of local interactions. 

Under these assumptions, a measure of the \textit{diversity} of the patches is a measure of sensitivity to initial conditions, in the sense of Lyapunov exponents. We are thus able to estimate dynamical complexity from a single SEM image, provided satisfying definitions of \textit{pattern} and \textit{complexity} are given.

\paragraph{(ii) Complex: diverse, but not too much} In ~\cite{lopez1995statistical}, the authors propose a novel statistical measure of complexity based on the following insight: physical systems that are complex are somewhere between an ideal gas, which is perfectly disordered, and a crystal, which is perfectly ordered. A good statistical measure of complexity can thus be defined as a product of a statistical measure of order and a statistical measure of disorder.

Shannon entropy\cite{shannon2001mathematical} is the most common measure of statistical disorder. Let $X$ be a system which can be in each of its $N$ states with probability $P(X=i)$, which we denote $P(X=i):=p_i$ when there is no risk of ambiguity and collect in a probability mass function $p:=\left\{p_i\right\}_{i=1\ldots N}$. Then Shannon entropy\cite{shannon2001mathematical}, which we denote $\mathscr{H}(X)$ or $\mathscr{H}(p)$ when there is no risk of ambiguity, is defined as:
\begin{align}
    \label{eq:shannon-entropy}
    \mathscr{H}(p) = - K\sum_{i=1}^N p_i \log p_i ,
\end{align}
where $K$ is a constant that effectively sets the base of the logarithm. In the sequel, we set $K=1$ and measure entropy in nats. Shannon entropy was introduced in the context of Information Theory as a measure of expected surprise when sampling from a random variable $X$ with probability mass function $p$. It has since seen powerful applications in a wide variety of fields ranging from Physics to Machine Learning to Biology. Shannon entropy is continuous in $p$, maximal for the uniform distribution where $\mathscr{H}(u) = \log N$, independent of zero-probability states. Finally, for any two discrete random variables $X,Y$ we have $\mathscr{H}(XY) = \mathscr{H}(X) + \sum_j P(Y=j) \mathscr{H}(X|Y=j)$, where $\mathscr{H}(X,Y)$ and $\mathscr{H}(X|Y=j)$ are, respectively, the entropy of the joint random variable $X,Y$ and the entropy of the random variable $X$ conditional on the state of $Y$ being $j$.

As Shannon himself remarked in \cite{shannon2001mathematical}, these properties are natural, in the sense they correspond intuitively to what one would expect from a measure of statistical diversity. Moreover, it can be shown that these four properties, known as the Shannon-Khinchin axioms~\cite{khinchin2013mathematical}, uniquely specify entropy, which justifies the Shannon entropy's standing as the de facto canonical measure of statistical diversity.

As to measures of order, we note that any decreasing function of $\mathscr{H}$, e.g. $(1-\mathscr{H})$ can be seen as a measure of \textit{order}. A trivial example is therefore $\mathscr{H}(1-\mathscr{H})$, which is simply a function of $\mathscr{H}$.

Since there is a canonical choice for the measure of disorder, the choice of order functional becomes crucial: among other things, it effectively sets the distributions that maximize complexity, which carries important semantic meaning. 

In~\cite{lopez1995statistical} the authors propose \textit{disequilibrium}, defined as the distance to equiprobability,  as a measure of order of $X$:
\begin{align}
\label{eq:disequilibrium}
    \mathscr{D} = \sum_{i=1}^N\left( p_i - \frac{1}{N}\right)^2,
\end{align}
The authors motivate this measure by expanding $H$ at $p$, a perturbation of magnitude $\delta p$ of the uniform distribution, which maximizes $\mathscr{H}$:
\begin{align*}
    \mathscr{H}(p) = \mathscr{H}_{max} -\frac{N}{2} \sum_{i=1}^N\left(\delta p_i - \frac{1}{N}\right)^2 + o\left(\|\delta p\|^2\right).
\end{align*}
We note that this implies that, up to a constant $D=1-\mathscr{H}(p) + o\left(\|\delta p\|^2\right)$: that is, disequilibrium coincides with the trivial choice of order functional up to second order in the expansion. Combining definitions \ref{eq:shannon-entropy} and \ref{eq:disequilibrium}, we define López-Ruiz - Mancini - Calbet complexity (\textit{LMC Complexity}):
\begin{align}
    \label{eq:lmc-complexity}
    C = \mathscr{H} \times\mathscr{D}.
\end{align}
We choose this measure in this report due to its intuitive appeal and ease of computation, as well as the wide range of applications, from image processing to fundamental physics and biomedical applications, that it has seen since its introduction\cite{lopezruiz2010statisticalmeasurecomplexity}.

\paragraph{(iii) Taylor complexity}

%The challenge of defining complexity lies in its inherently subjective nature, as there is no single, universally accepted measure of it.

How can we measure the diversity of shapes at a given point $x$? Take, for example, the shapes that we obtain from two real functions, $f(x) = a_1x^2$ and $g(x) = sin(a_2x)-sin(\sqrt{2}a_2x)$, where $a_1, a_2\in \mathbb R$ are sampled from the same discrete random variable $A$. Intuitively, the set of shapes that we obtain in the first case is a lot less diverse than the second one, even thought their source of diversity, $A$, is the same. 

A variance-type measure of diversity of shapes could be obtained by adding the variances of the shape components, since for independent random variables $X$ and $Y$:
\begin{align}
\label{eq:5}
\textrm{Var}(X+Y) = \textrm{Var}(X) + \textrm{Var}(Y).
\end{align}

If $x$ is small, in our $f$ and $g$ example above, the greatest contribution to 
the diversity of the two shapes comes from the $g$ part. To see this, note that since $\sin x\sim x$ as $x\rightarrow 0$, we can approximate the variance as $\mathrm{Var}(xA)-\mathrm{Var}(\sqrt{2}xA) = (1-\sqrt{2})^2x^2\mathrm{Var}(A)$, which is much greater than $\mathrm{Var}(Ax^2) = x^4\mathrm{Var}(A)$ when $x$ is small. Conversely, if $x$ is large, then most of the diversity comes from the $f$ part. 

With entropy as a measure of diversity, again for independent $X$ and $Y$, we obtain (applying the third Shannon-Khinchin axiom above to $\mathscr{H}(X+Y,Y)$, and noting that for independent $X,Y$ we have $\mathscr{H}(X+Y|Y) = \mathscr{H}(Y)$) :
\begin{align}
\label{eq:7}
\mathscr{H}(X+Y) = \mathscr{H}(X) + \mathscr{H}(Y) - \mathscr{H}(Y|X+Y)
\end{align}
The last term $\mathscr{H}(Y|X+Y)$ is zero only if each sum $x+y\sim X+Y$ \textit{uniquely} determines $y\sim Y$, which is generally not the case. Moreover, the diversity as measured by the entropy no longer has a clean dependency on the magnitude of $x$ which means that we can no longer neatly divide the contributions of each of the parts to the entropy of $f+g$. 

Hence, in general, the two measures of diversity thus correspond to different ways to combine component diversity. Similar arguments could be given for other statistics, which may have very different ways to be expressed in terms of statistics of given decompositions.

We can avoid these ambiguities by examining the decomposition of shape at a point $x+h$ into components that change at different rates with $h$, the distance to $x$, specifically as a sum of terms proportional to powers of the distance $h$. Setting $u:\mathbb R\rightarrow \mathbb R$ for simplicity, we have:
\begin{align}
\label{eq:1}
u(x+h) = O(h^0) + O(h^1) + O(h^2) + \ldots
\end{align}
When $h$ is small and all constants multiplying the powers of $h$ are of the same magnitude,  most of the contribution to $u(x+h)$ comes from the first term $O(h^0)$; higher-order terms do not contribute much to the actual value of $u(x+t)$. To fix intuition, setting $h=0.1$, one can think of the constant multiplying $h^n$ as the $n$-th decimal digit in a decimal expansion, which becomes increasingly negligible for larger $n$.

Moreover, the provided $u$ is sufficiently well-behaved since Taylor's Theorem allows us to compute the digits in this expansion using Taylor's expansion:
\begin{align}
\label{eq:2}
u(x+h) = \sum\limits_{i=0}^{\infty} \frac{u^{(n)}(x)}{n!}h^{n}
\end{align}
Conversely, if $h$ is large and all constants are of the same magnitude, then the least significant contribution to $u(x+h)$ comes from the first term.

In a physical setting "small" and "large" are given in terms of a length that is, in some way, fundamental to the problem: the \textit{scale}. This brings us to the case where $h$ is neither large nor small, that is, of order unity, $h=1$, measured in that fundamental scale (specifically, in an image,  fundamental length is given by the inter-pixel distance).

At the distance of one fundamental length, Taylor's formula thus reads:
\begin{align}
\label{eq:3}
u(x+1) =  u(x) + u^{(1)}(x) +\frac{u^{(2)}(x)}{2!} + \ldots,
\end{align}
which tells us how to weight each of the derivatives in order to compute $u$ at a distance of 1 fundamental length from $x$. In other words, if we consider the value of the function and its derivatives at $x$, each contributes to the ``shape'' of the field at that scale, with higher-order derivatives being weighted by $1/n!$.  

We use the reasoning above to introduce a weighting for shape statistics of a field in terms of the values of the field and $k$ of its derivatives: that given by the terms of the Taylor expansion.

Specifically, given a statistic $S$ of a field $u$ (e.g. mean, variance, ...), we define its associated \textit{Taylor-n statistic} as:
\begin{align}
\label{eq:taylor-statistics}
T^n S(u) = \sum\limits_{k=0}^{n} \frac{S(u^{(k)})}{k!}.
\end{align}
Here, $u^{(k)}$ denotes the $k$-th derivative of $u$. If $u$ is multi-dimensional (e.g.\ an image), an analogous expression can be written in terms of partial derivatives. Specifically, for $u:\mathbb R^p\rightarrow \mathbb R$ we have:
\begin{align}
\label{eq:taylor-statistics-multidim}
T^n S(u) = \sum\limits_{|\alpha|=0}^{n} \frac{S(u^{(\alpha)})}{\alpha!},
\end{align}
where $\alpha$ is a multi-index $\alpha =(\alpha _{1},\alpha _{2},\ldots ,\alpha _{n})$ of length $n$, with $|\alpha |\ =\ \sum _{k=1}^{n}\alpha _{k}$, the factorial $\alpha !$ denoting $\alpha _{1}!\cdot \alpha _{2}!\cdots \alpha _{n}!$ and $u^{(\alpha)}:=\frac{\partial^{\alpha_1+\cdots \alpha_n}}{\partial x_1^{\alpha_1}\cdots \partial x_n^{\alpha_n}}u$.

We thus obtain a statistic that incorporates the value and the local shape, by weighting $u$ and its spatial derivatives in the same proportion that they contribute to the change in $u$ at a distance of one fundamental length from $x$, (namely $1/k!$).

% For a given field $u(\mathbf{x}, t)$ the diversity of patches is quantified using Shannon entropy:
% \begin{align}
%   H= -K\sum_{i=0}^{N} p_{i} \log p_{i}  
% \end{align}
% where $N$ is the total number of patterns and $p_{i}$ is the distribution of sign patterns. As the derivatives represent terms in the Taylor expansion and are independent (per Taylor's theorem), the overall diversity of the patterns is calculated by summing the diversities of the individual derivatives, weighted by their respective coefficients in the Taylor expansion:
% \begin{align}
%     \mathscr{H}(u) = \mathscr{H}(|u|) + H\left(\left|\frac{\partial u}{\partial x}\right|\right) + H\left(\left|\frac{\partial u}{\partial y}\right|\right) \notag  \\ + \frac{1}{2} H\left(\left|\frac{\partial^{2} u}{\partial x^{2}}\right|\right) + H\left(\left|\frac{\partial^{2} u}{\partial x \partial y}\right|\right) + \frac{1}{2}H\left(\left|\frac{\partial^{2} u}{\partial y^{2}}\right|\right)
% \end{align}

% This "information" is completed by the measure of the "disequilibrium" a measure of a probabilistic hierarchy \cite{lopez1995statistical}. The disequilibrium $\mathscr{D}$ is the quadratic distance of the actual probability distribution $p_{i}$ to equiprobability and is expressed :
% \begin{align}
%     \mathscr{D}(u)=\sum_{i=0}^{N}\left(p_{i}-\frac{1}{N}\right)^{2}
% \end{align}

% Finally, the complexity is measured with LMC complexity \cite{catalan2002features} and written :
% \begin{align}
%     \mathscr{C}(u) = \mathscr{H}(u)\mathscr{D}(u)
% \end{align}

With a satisfiable definition of pattern-statistics, we turn to the problem of computational efficiency. To do so, we turn to \textit{permutation entropy} of Bandt and Pope\cite{bandt02_permut_entrop}, which we briefly recall. The problem is similar to ours, that of estimating dynamic complexity, but in the time-series setting. This can be done via traditional measures such as Lyapunov exponents or Kolmogorov-Sinai entropy, which are unfortunately intractable to compute and sensible to noise. To address these difficulties, the authors propose computing the Shannon entropy of \textit{ordinal patterns} instead. Specifically, the ordinal pattern of length $k$ at $u(t)$ is the permutation that arranges $u(t), u(t+1), u(t+2),\ldots, u(t+k-1)$ in increasing order. This measure is very simple and efficient to compute. Moreover, it can be shown that, in certain situations, it coincides with Kolmogorov-Sinai entropy, which is a measure of dynamical diversity in the sense of chaos, and which is generally intractable. Intuitively, this is possible because we replace the full time-series dynamics with a "collapsed" version that tracks local pattern dynamics. This collapse retains enough dynamic information to compute dynamic diversity,  but is important enough that computation becomes possible.

Applying this measure to higher dimensions is not straightforward because there is no natural order in dimension $n\geq 1$. This motivates a more general definition of "pattern" that is applicable to higher dimensions. Suppose that the field $u$ has been sampled at a number of grid points. Then the \textit{sign-pattern field} associated to $u$ is the mapping that assigns to each grid point $x$ a binary vector of dimension equal to the number of neighbors of $x$ and values $1$ if $u$ computed at that location is greater than the mean value of $u$ and zero otherwise (the order of the components is arbitrary but fixed for all $x$). Specifically for a square grid in two dimensions, the sign pattern field of $u$ at $x_{ij}$, which we denote by $Pu_{i,j}$ is the 9-component vector:
\begin{align*}
Pu_{i,j} = (&u_{i-1,j-1}>0, &u_{i-1,j}>0, &u_{i-1,j+1}>0,\\
            &u_{i,j}>0,     &u_{i,j-1}>0, &u_{i,j+1}>0,\\
            &u_{i+1,j-1}>0, &u_{i+1,j}>0, &u_{i+1,j+1}>0). 
\end{align*}
Note that one can also see $Pu_{ij}$ as a 9-bit binary number, and that computing a statistic on $Pu$, seen as a collection of independent samples, is straightforward.

Computing the sign patterns of $u$ and its spatial derivatives up to order $2$, we can use the 2-dimensional version of~\ref{eq:taylor-statistics}, with $n=2$ to produce statistics that are local in the sense explained in the discussion above. Specifically, for the LMC complexity $C$ defined in~\ref{eq:lmc-complexity} we have:
\begin{align*}
    T^2C(Pu) = C(Pu) + C(Pu_x) + C(Pu_y)\\ + \frac{1}{2}C(Pu_{xx}) +C(Pu_{xy}) + \frac{1}{2}C(Pu_{yy}),
\end{align*}
where in our specific setting, the derivative fields are all finite difference approximations of order 2 using 5-point stencils, which ensures that the information in each sign pattern is consistent among all derivative fields.

We call this measure of complexity \textit{Taylor complexity} (or simply \textit{complexity} when there is no risk of ambiguity), and use it in all our computations.

% To quantify the diversity of these patches while accounting for spatial structure, we examine the spatial derivatives present in $f$. Specifically, if $f(x_{i}, . . . , \partial_{i}^{j} , . . .)$ includes derivatives, we define the smallest patch size $k$ that enables computation of all such derivatives using a finite difference scheme. 

% This patch size represents the locality of $f$. With this patch size we have enough information to compute the 2nd  order in space derivative fields of up to order 2 approximation. It means that the patch size is good to reconstruct the field as a Taylor series up to the second order terms. This defines a notion of locality: the radius that allows this reconstruction of the original field. Since each patch corresponds to a distinct evolution of the system from perturbed initial conditions, the diversity of patches reflects the sensitivity to these conditions and can be seen as a measure of the complexity of $f$. 

% Then, we use the distribution of sign patterns $\{ -, 0, +\}$ of the derivatives compatible with the patch size. For example, a $3\times3$ patch (suitable for derivatives up to second order) has $3^{9}$ possible sign patterns for each derivative. But in reality, since the number of points where the derivative is zero is small, and highly dependent on noise, we randomly set the $0$ value to $+$ or $-$ with equal probability. This corresponds to $2^{9}$ possible sign patterns for each derivative. 

\begin{figure*}[!htbp]
    \begin{tikzpicture}
        \node (bottle1) {\includegraphics[width=0.97\textwidth, keepaspectratio]{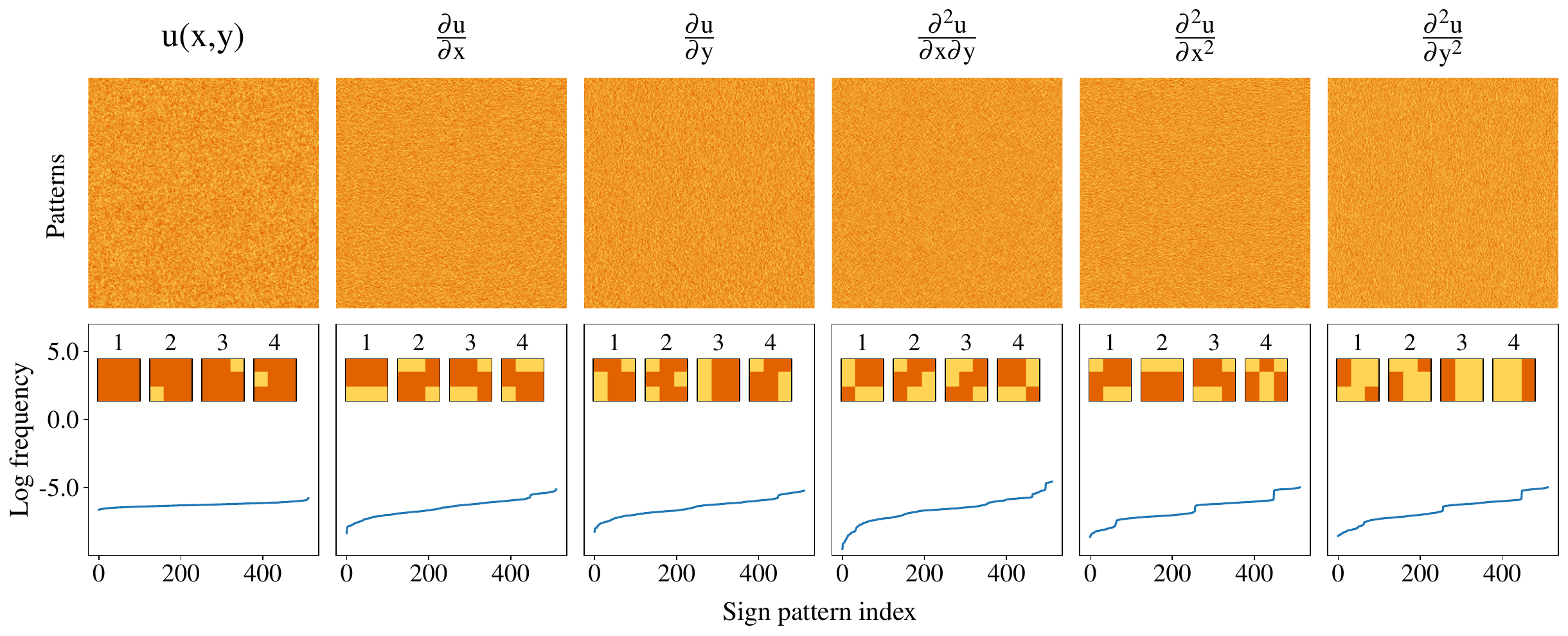}};
        \node [left=-0.3cm of bottle1.north west, yshift=-4.0mm, xshift=-2.0mm] {\textbf{a}};
        
    \end{tikzpicture}
    \vspace{-0.3cm} 

    \begin{tikzpicture}
        \node (bottle2) {\includegraphics[width=0.97\textwidth, keepaspectratio]{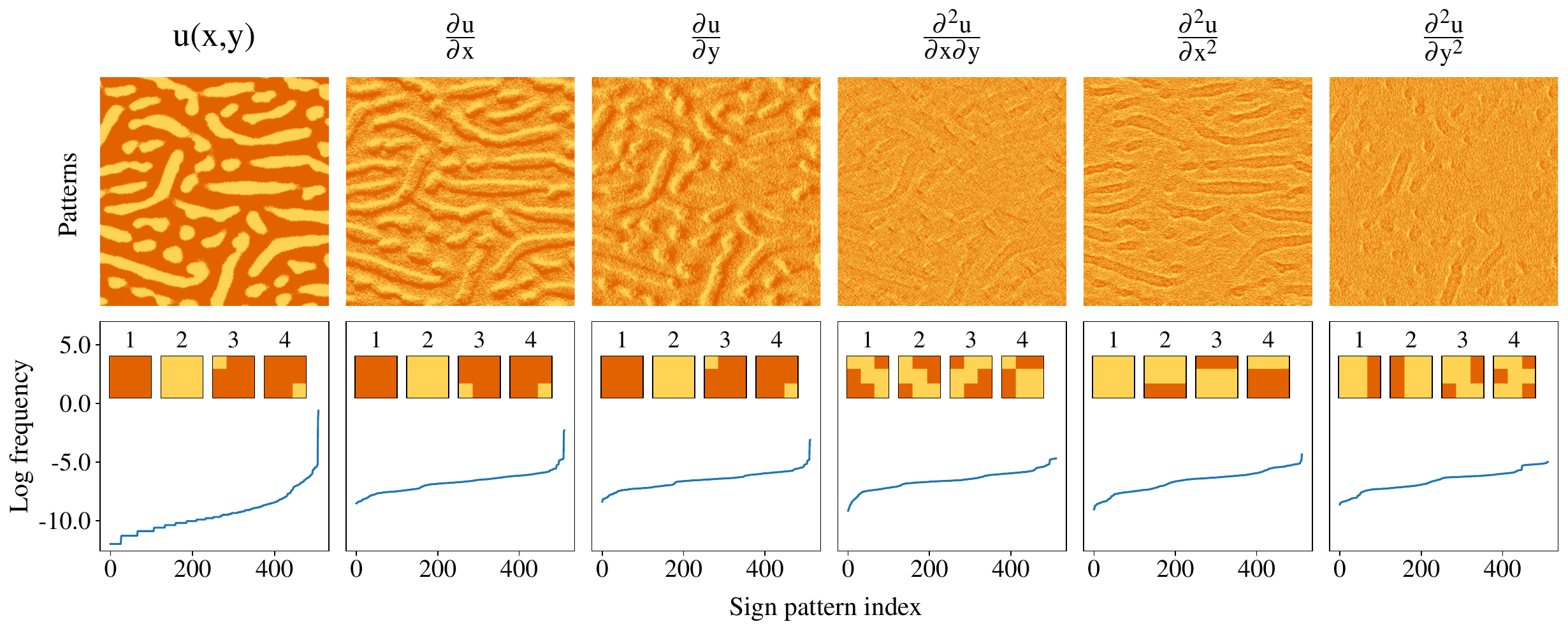}};
        \node [left=-0.3cm of bottle2.north west, yshift=-5.5mm, xshift=-2.0mm] {\textbf{b}};
        
    \end{tikzpicture}

    \begin{tikzpicture}
        \node (bottle3) {\includegraphics[width=0.97\textwidth, keepaspectratio]{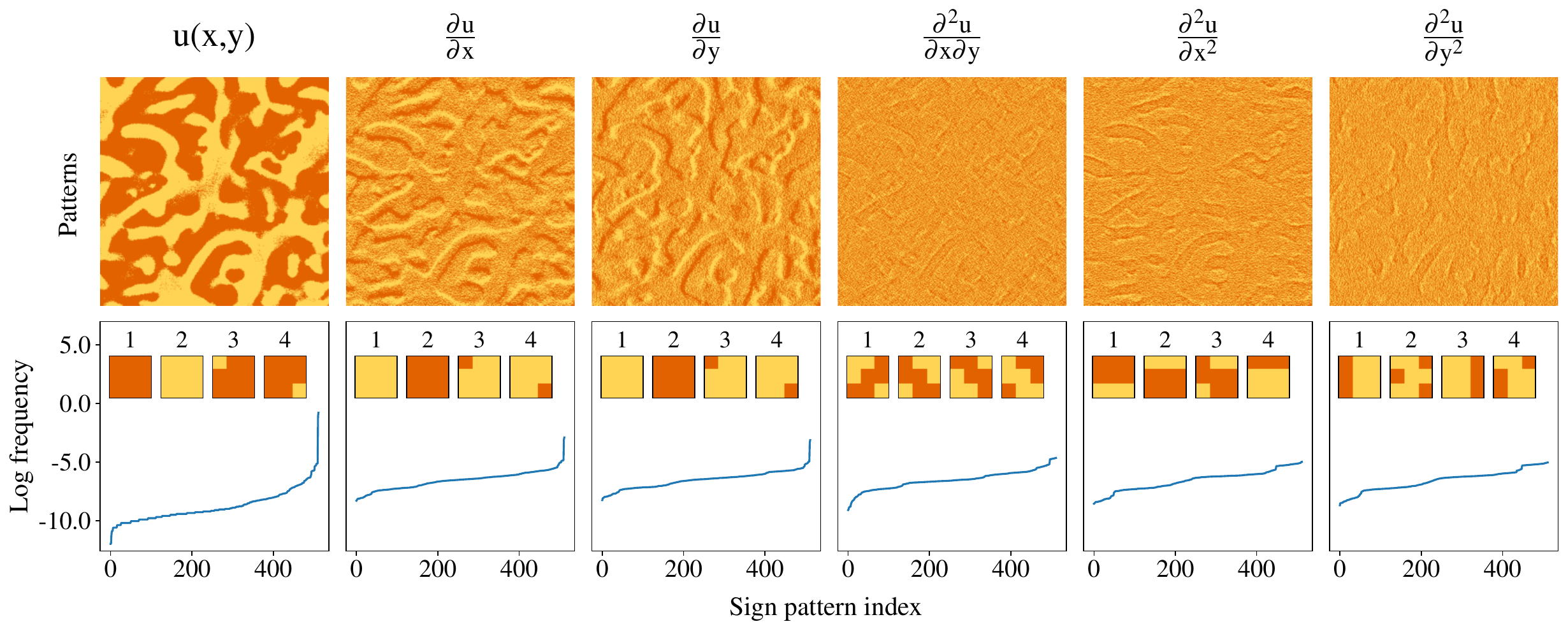}};
        \node [left=-0.3cm of bottle3.north west, yshift=-5.5mm, xshift=-2.0mm] {\textbf{c}};
        
    \end{tikzpicture}
\captionsetup{justification=justified, singlelinecheck=false}     
\caption{\textbf{Illustration of complexity computation considering local information of a surface.} The plot represents three different patterns: (a) a flat pattern, (b) an organized pattern, and (c) a chaotic pattern. In each subfigure, the top row displays the sign of the indicated field after subtracting the mean: yellow for pixel values $\geq$ 0, orange for pixel values $<$ 0. The second row shows the $3\times3$ sign distribution of the fields above. The small squares are the 4 most frequent patterns, with orange indicating pixel values $<$ 0 and yellow indicating pixel values $\geq$ 0.} \label{fig:figure_complex}
\end{figure*}

Fig.~\ref{fig:figure_complex} illustrates how surface complexity is quantified by analyzing the sign distribution of a field $u(x,y)$ and its partial derivatives up to the second order. The three subfigures represent a flat surface, an organized pattern, and a chaotic pattern. The first row shows the sign distribution after subtracting the mean. The second row depicts the frequency of occurrence of a $3 \times 3$ patch in the field $u(x,y)$, sorted in ascending order by a sign pattern index which ranks them from the least to the most frequent. There are $2^{9}$ possible sign patterns for each field, without any inherent order. The ascending sorting of patterns is arbitrary and serves only to systematically organize their frequency of occurrence. The small inset square highlights the four most frequent patterns, displayed in descending order from left to right.

\section{Artificial images generation process}
In order to maximize absorbed energy, the matter self-organizes, leading to variations in its complexity $\mathscr{C}$ to produce an optimal pattern $\mathscr{P}$. This suggests that the emergence of patterns is not random but a consequence of an optimization process driven by energy absorption. To test this hypothesis, we study the effect of perturbing $\mathscr{P}$ on absorbed energy by generating artificial images. Our expectation is that arbitrary modifications of $\mathscr{P}$, which initially maximize $E_{abs}$, will disrupt the pattern's efficiency and lead to a reduction in absorbed energy. 

Formally, we express this relationship as:
\begin{align}
   E_{abs} = E_{abs}(\mathscr{C}(\mathscr{P})),  
\end{align}

which, upon differentiation, gives:
\begin{align}
 \frac{dE_{abs}}{d\mathscr{P}} = \frac{dE_{abs}}{d\mathscr{C}} \cdot \frac{d\mathscr{C}}{d\mathscr{P}}  \label{eq:eabs_derivative}
\end{align}

This formulation implies that any change in the pattern $\mathscr{P}$ influences absorbed energy $E_{abs}$ indirectly through its effect on complexity $\mathscr{C}$. If a perturbation in $\mathscr{P}$ does not alter $\mathscr{C}$, $\frac{d\mathscr{C}}{d\mathscr{P}} = 0$, then the derivative of absorbed energy with respect to $\mathscr{P}$ remains zero, $\frac{dE_{abs}}{d\mathscr{P}} = 0$, meaning absorbed energy remains unchanged. 

On one hand, if the pattern $\mathscr{P}$ is modified in a way that reduces complexity, $\frac{d\mathscr{C}}{d\mathscr{P}} \leq 0$, as we replace patterns with lower complexity ones, the resulting decrease in complexity $\mathscr{C}$ leads to a decline in absorbed energy $E_{abs}$, implying that $\frac{dE_{abs}}{d\mathscr{C}} > 0$.

On the other hand, if the pattern $\mathscr{P}$ is altered by a physical process that tends to increase complexity, $\frac{d\mathscr{C}}{d\mathscr{P}} \geq 0$, while maintaining the absorbed energy constant, $\frac{dE_{abs}}{d\mathscr{P}} = 0$, by applying the product rule of differentiation we have $\frac{d\mathscr{C}}{d\mathscr{P}} = 0$. This confirms that the observed patterns correspond to local maxima of complexity.

\subsection{Swift-Hohenberg method}

To assert this constrained maximization statement, we thus perturb the original field $\mathscr{P}$ using perturbations $\delta u_{SH}$ that are compatible with the physical process. Specifically, one takes a small patch $P$ of the original field $\mathscr{P}$ and solves the Swift-Hohenberg equation $SH(r, \gamma, \mathbf{q}_0)$, where $r$ is the bifurcation parameter, $\gamma$ introduces asymmetry by breaking the symmetry of the solutions with respect to sign inversion, and $\mathbf{q}_0$ is the typical spatial wavenumber. This equation is applied to $P$ for the initial condition:
\begin{align}
  u(\mathbf{x}, 0) = \mathscr{P}|_P(\mathbf{x})
\end{align}

with Neumann boundary conditions:
\begin{align}
    u|_{\partial P}(\mathbf{x}, t) = \mathscr{P}|_{\partial P}(\mathbf{x})
\end{align}

We use the pseudospectral solver developed for a previous work \cite{brandao22_learn_pde_to_model_self_organ_matter}, sampling $r \in (0,1]$ and $\gamma \in [-1,1]$, which allows the generation of a good variety of pattern solutions. These values of $r$ and $\gamma$ help control the nature of the perturbations, allowing for a broad exploration of the solution space that remains consistent with the physical constraints imposed by the field.

This ansatz guarantees that the generated perturbations will be compatible with the physical process. The "amount" of perturbation is equated to the area of the field, while the specific shape of the boundary is considered unphysical. To control possible artifacts, we minimize the boundary length. The shape that minimizes the perimeter for a given area is a circle. Unfortunately, this shape is highly symmetrical and will induce radial symmetry in the solutions. Hence, we choose to set the shape to \emph{locally} minimize the perimeter and set the shape as a convex domain. We sample the convex domain using the Valtr algorithm \cite{valtr95_probab_that_random_point_are_convex_posit} to generate a 100-vertices convex polygon (the algorithm samples uniformly from the 100-vertices convex polygon).

Finally, to control size artifacts, we set $\mathbf{q}_{0}$ using the modal spatial frequency in the radially averaged power spectrum.

To summarize, the process for generating physical perturbations $\mathscr{P}+\delta \mathscr{P}$ is outlined as follows:

\begin{enumerate}
\item Given field $\mathscr{P}$, sample uniformly a center $c$ for the domain location.
\item Sample a patch $P$ of the initial field $\mathscr{P}$, a 100-vertices convex polygon uniformly using the Valtr algorithm.
\item Set $\mathbf{q}_{0}$ using the modal spatial frequency in the radially-averaged power spectrum in $P$.
\item Sample $r \sim U(0,1]$ and $\gamma \sim U([-1, 1])$.
\item Solve $SH(r, \gamma, \mathbf{q}_0)$ on $P$ with the initial condition $u(\mathbf{x}, 0) = \mathscr{P}|_P(\mathbf{x})$ and Neumann boundary conditions $u|_{\partial P}(\mathbf{x}, t) = \mathscr{P}|_{\partial P}(\mathbf{x})$ for a fixed (short) evolution time $t$.
\item Replace the original $P$ with the SH-transformed $P$.
\end{enumerate}

By following these steps, we generate a diverse set of physical perturbations of the original image, each governed by the Swift-Hohenberg equation. The resulting patterns are denoted $\mathscr{P}_{SH}$ in the main text.

\subsection{Random patch method}

In this method, we perturb the original field $\mathscr{P}$ using a perturbation that is incompatible with the physical process of self-organization. We begin by considering a target pattern corresponding to the field $\mathscr{P}$ and a source pattern $\mathscr{S}$ coming from the learning regime. The idea is to replace a small patch $P$ from the target field $\mathscr{P}$ with a corresponding patch $S$ from the source field $\mathscr{S}$. Both $P$ and $S$ are chosen from convex domains, which are generated using the same Valtr algorithm previously applied in the Swift-Hohenberg method. The Valtr algorithm ensures that both patches $P$ and $S$ are convex polygons with 100 vertices, sampled uniformly. 

The perturbation process can be summarized in the following steps:

\begin{enumerate}
    \item Given the source field $\mathscr{S}$ and the target field $\mathscr{P}$ fields, sample uniform centers, $c_s$ and $c_p$ for domain locations in each field
    \item Sample a patch $P$ of the target field $\mathscr{P}$ and a patch $S$ of the source field $\mathscr{S}$. $P$ and $S$ are identical, both being 100-vertices convex polygons sampled uniformly using the Valtr algorithm
    \item Replace the target field patch $P$ with the source field patch $S$
\end{enumerate}

These steps generate diverse non-physical perturbations to the original image, introducing varied features to analyze system dynamics under perturbation. In the main text, the resulting patterns are represented as $\mathscr{P}_{RP}$.

\begin{figure*}[!htbp]
\centering
\includegraphics[width=1\textwidth]{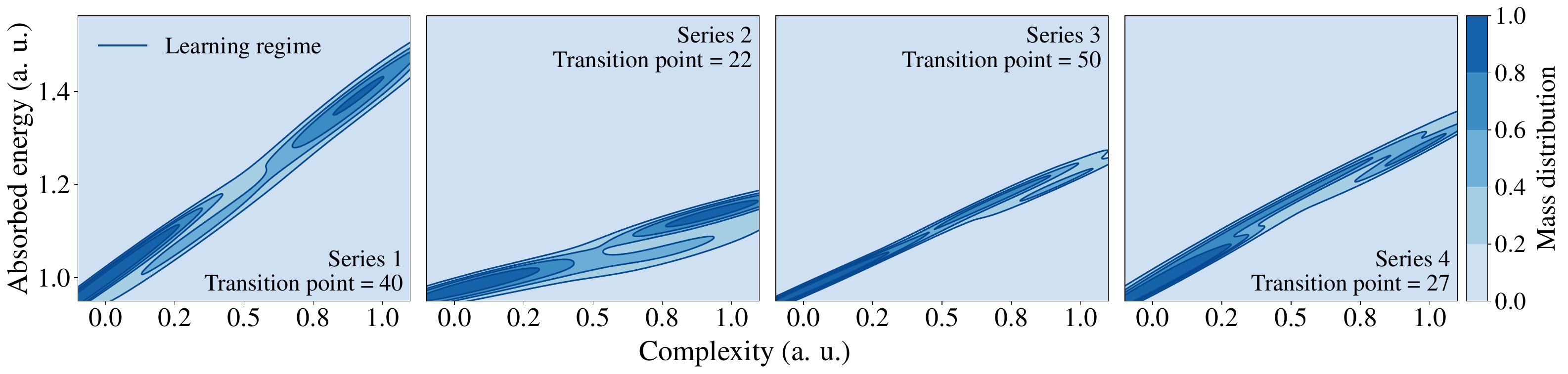}
\captionsetup{justification=justified, singlelinecheck=false} 
\caption{\textbf{Estimation of data distribution during learning regime}. Each column represents different series, with the transition point indicating the number of data points used in distribution estimation. The contour divides the distribution into five regions, covering mass ranges from $0-20\%$ to $80-100\%$.}
\label{fig:figure_mutual_info}
\end{figure*}

\section{Estimation of distribution during the learning regime}
To understand how the system evolves before the emergence of chaotic structures, we analyze the data distribution during the learning regime, focusing on the relationship between absorbed energy $E_{abs}$ and complexity $\mathscr{C}$.

As the number of double pulses $N$ increases, the system transitions from the learning to the chaotic regime. We define the transition point as the value of $N$, that marks the beginning of this shift. To estimate the distribution during the learning regime using the kernel density estimation (KDE), we consider data points up to the transition point $N$, meaning we incorporate all data points $x_{1}$, $x_{2}$, ... $x_{N}$ where $x_{N}=[E_{abs}^{N}, \mathscr{C}^{N}]^T$ with \( E_{abs}^N \) and \( \mathscr{C}^N \) being respectively the absorbed energy and complexity after $N$ double pulses. For an intuitive representation, we adjust the contours and colormap to reflect mass distribution. As shown in Fig.~\ref{fig:figure_mutual_info}, the mass is partitioned into five regions, ranging from $0-20\%$ to $80 - 100\%$, with colors transitioning from light blue to dark blue. The four series exhibit a comparable distribution, characterized by contours that reveal a mass spread primarily along a diagonal axis. The distribution is broader and less concentrated in any specific region. 

This suggests a positive correlation between \( E_{abs} \) and \( \mathscr{C} \) during the learning regime. As complexity increases, so does self-organization, and a higher complexity value signifies a more structured pattern. It indicates that the material reorganizes to optimize energy absorption. Therefore, energy absorption is not a random process but follows an underlying mechanism that promotes organization before chaos emerges.

\section*{Methods}
\subsection*{Kernel density estimation}
Kernel density estimation (KDE) is a popular method for estimating a dataset’s probability density function without assuming a predefined parametric model \cite{parzen1962estimation}. KDE learns the distribution directly from the data, making it well-suited for capturing complex patterns. The fundamental principle of KDE is to use a kernel function to generate a smooth density estimate.

The density $\hat{f}(\mathbf{x})$ at a point $\mathbf{x}=[E_{abs}, \mathscr{C}]^T$ is given by:
\begin{align}
    \displaystyle\hat{f}(\mathbf{x}) = \frac{1}{M} \sum_{N=1}^M K \left(\mathbf{x} - x_N\right),
\end{align}

where $\mathbf{x}$ is the point at which we want to estimate the density, $x_{N}=[E_{abs}^{N}, \mathscr{C}^{N}]^T$ is an individual data point with \( E_{abs}^N \) and \( \mathscr{C}^N \) being respectively the absorbed energy and complexity after $N$ double pulses, $M$ is the total number of data points in the dataset and $K(\mathbf{x})$ is a bivariate gaussian kernel function which defines the shape of the smoothing, expressed as: 
\begin{align}
   K(\mathbf{x})=\frac{1}{2\pi \sqrt{det(H)}} \exp{(\mathbf{x}^{T}H^{-1}\mathbf{x})}
\end{align}

where $H$ is $2\times2$ matrix that governs the form of the function.

\bibliographystyle{abbrv}
\bibliography{biblio}